\documentclass{aa}

\usepackage{hyperref}
\usepackage{pdflscape}
\usepackage{xcolor}
\usepackage{lscape}
\usepackage{float}
\usepackage{balance}
\usepackage{soul}

\usepackage{graphicx}

\def\M$_\odot${M$_{\odot}$}

\def\Msol{M$_\odot$}
\def\M$_\odot${M$_\odot$}

\usepackage{txfonts}

\begin{document}

   \title{Southern massive stars at high angular resolution}
   \subtitle{WR~25 is a massive hierarchical triple system\thanks{Based on observations collected at the European Southern Observatory (ESO) under ESO program ID 0102.D-0234 (PI: Sana).}}

   \author{K.\ Deshmukh\inst{\ref{inst:kul},\ref{inst:lgi},}\thanks{Corresponding author; \texttt{astro.kunal.deshmukh@gmail.com}}
          \and L.\ Mahy\inst{\ref{inst:rob}}
          \and H.\ Sana\inst{\ref{inst:kul},\ref{inst:lgi}}
          \and A. J.\ Frost\inst{\ref{inst:eso}}
          \and E.\ Gosset\inst{\ref{inst:liege}}
          \and C.\ Lanthermann\inst{\ref{inst:chara}}
          \and 
          J.-B.\ LeBouquin\inst{\ref{inst:gren}}
          \and \\
          D.\ Pauli\inst{\ref{inst:kul}}
          \and T.\ Shenar\inst{\ref{inst:tau}}
     }

   \institute{
{Institute of Astronomy, KU Leuven, Celestijnenlaan 200D, 3001 Leuven, Belgium\label{inst:kul}}
\and 
{Leuven Gravity Institute, KU Leuven, Celestijnenlaan 200D, box 2415 3001 Leuven, Belgium \label{inst:lgi}}
\and
{Royal Observatory of Belgium, Avenue Circulaire/Ringlaan 3, B-1180 Brussels, Belgium \label{inst:rob}}
\and
{Space sciences, Technologies and Astrophysics Research (STAR) Institute, Université de Liège, Allée du 6 Août, 19c, Bât. B5c, B-4000 Liège, Belgium \label{inst:liege}}
\and
{European Southern Observatory, Santiago, Chile \label{inst:eso}}
\and
{The CHARA Array of Georgia State University, Mount Wilson Observatory, Mount Wilson, CA 91203, USA \label{inst:chara}}
\and
{Institut de Plan\'etologie et d’Astrophysique de Grenoble, Grenoble 38058, France \label{inst:gren}}
\and
{The School of Physics and Astronomy, Tel Aviv University, Tel Aviv 6997801, Israel \label{inst:tau}}
}

   \date{Received -; Accepted -}



\abstract{WR~25 is a massive colliding-wind binary in the Carina nebula comprising a WN6ha primary star with an O5 companion in an eccentric 208-day orbit. Recent spectroscopic analysis estimates the total binary mass to approach $100\,M_\odot$, and a primary-to-secondary mass ratio of $q= M_1/M_2\approx2$, hinting toward a very massive primary star. The presence of additional spectroscopic signatures from a third, ``intruder'' star was also noted, making it a candidate hierarchical triple system.

In this study, we present a VLTI/PIONIER interferometric observation of WR~25, spatially resolving all three components for the first time.
For the inner WN6ha + O5 binary, we find a projected angular separation of $1.68\pm0.02$ milliarcseconds (mas), corresponding to a projected physical separation of $3.90\pm0.21$~au at a distance of $2.32 \pm 0.12$ kpc. Leveraging the fortunate timing of the VLTI observation, which was obtained when the two components were passing the line of nodes, we determined the orbital semi-major axis $a=3.11\pm0.20$~au. 
Subsequently, the newly constrained total dynamical binary mass is $93\pm18\,M_\odot$, with a primary mass $M_1=62\pm13\, M_\odot$ and secondary mass $M_2=31\pm7\, M_\odot$. 

We detect the third star or tertiary with a projected angular separation of $27.69\pm0.02$ mas from the primary, with a chance alignment probability lower than $10^{-4}$. Using newly obtained brightness ratios between all components, we revisit archival spectroscopic data of WR~25 to disentangle spectra for individual components and derive their stellar parameters. 
The tertiary, which has a spectral type O7, is broadly coeval with the inner binary and has an evolutionary mass $M_3=25.6^{+2.8}_{-2.3}\,M_\odot$.  Based on simulations, we estimate the tertiary period to be in the range 19 -- 82 years. 
The newly confirmed triple nature of WR~25 thus makes it an important benchmark system to measure accurate dynamical masses of the inner binary and potentially the tertiary, to calibrate stellar evolution and atmosphere models, and to study its formation and stability as a hierarchical triple system.
}


\titlerunning{WR~25 is a massive hierarchical triple}
\authorrunning{K. Deshmukh et al.}

\maketitle
%
%
\section{Introduction}
\label{sec:intro}

Massive stars ($M_{\rm ini}>8\,M_\odot$) immensely influence their surroundings through life and death. The earliest O-type stars with initial masses $M_{\rm ini}>60\,M_\odot$ can evolve to become extremely luminous (log($L/L_\odot) \gtrsim 5.8-6.0$) and drive strong winds even on the Main Sequence \citep{1994Langer, 2007Crowther,2008Smith}. Subsequently, their spectra are dominated by strong emission lines of hydrogen, helium and nitrogen, similar to nitrogen-rich Wolf-Rayet stars but with residual hydrogen (WNh). These stars release copious amounts of ionizing radiation and nuclear processed material into their surroundings throughout their few-Myr lifetime, and possibly also during their core collapse, leaving behind a stellar-mass black hole \citep{2012Langer,2019Woosley,2022Eldridge}. They are hence rare and crucial laboratories for some of the most extreme physical phenomena and warrant special attention.

Almost all O-type stars are formed and live in multiple systems \citep{2012Sana,2014Sana,2023Lanthermann}, which plays a key role in their evolution, e.g. in the form of tidal effects, mass transfer, common envelope evolution and mergers \citep{2013deMink}. In addition, massive binary \citep{2018Kruckow,2024Marchant} and triple \citep{2026Stegmann} evolution are thought to be among the leading formation channels for merging stellar-mass compact objects detected with LIGO-Virgo-Kagra \citep[e.g.][]{2025gwtc}. Massive stars in multiple systems also offer a special opportunity to constrain one of their most fundamental properties - their masses.
In a broader context, measuring masses is an important step toward calibrating our knowledge of stellar structure and evolution.

Spectroscopic binaries that are also eclipsing allow for accurate dynamical mass measurements. Only a handful of such binaries are known that host massive WNh stars, such as: WR 20a \citep[$M_1 =83\pm5\,M_\odot$ and $M_2=82\pm5\,M_\odot$,][]{2004Rauw,2004Bonanos}, WR 21a \citep[$M_1=93\pm2\,M_\odot$ and $M_2=53\pm1\,M_\odot$,][]{2016Tramper,2022Barba}, NGC 3603-A1 \citep[$M_1=93\pm11\,M_\odot$ and $M_2=70\pm9\,M_\odot$,][]{2025Massey}. Because of the need to have eclipses, these studies tend to be limited to short periods ($P\lesssim30\,{\rm days}$). In some cases, binary component masses can also be constrained using wind eclipses or polarization, although both methods rely on modeling the wind of the WNh stars \citep{1988StLouis,1996Lamontagne,2017Shenar,2021Shenar}.
For longer periods, and thus larger orbital separations, an alternative way to measure dynamical masses is using optical long-baseline interferometry to obtain a relative astrometric orbit \citep{2017LeBouquin}. With complementary information about the distance (e.g. from {\it Gaia}) and mass ratio from spectroscopic measurements, a full 3-dimensional (3D) orbit and accurate model-independent masses can be obtained \citep[e.g.][]{2024Richardson,2024Holdsworth}. WR~89 is the only massive WNh star so far with a spatially resolved companion detected with optical interferometry, pending follow-up observations for a full orbit determination \citep{2024Deshmukh}.

WR~25 or HD~93162 was first identified as a massive eccentric WNh + O binary system with a 208-day orbital period by \citet{2006Gamen}. Residing in the Carina nebula, WR~25 is a member of the Trumpler 16 cluster \citep[distance $D=2.32\pm0.12$\,kpc,][]{2021Shull} and lives among some of the youngest and most massive stars known in the Galaxy. It has been a well-documented bright source in X-ray studies of the region for decades \citep{2003Raassen,2008Antokhin,2024Sasaki}. More focused studies have also extensively characterized and established WR~25 as a colliding-wind binary system \citep{2006Pollock,2014Pandey,2019Arora,2021Pradhan}. A detailed spectroscopic study was recently presented by \citet{2026Gosset}, which reclassified the system as a WN6ha + O5 binary and updated its orbital parameters, also estimating a total binary mass approaching $100\,M_\odot$. The updated orbital parameters are listed in Table\,\ref{tab:orb}. In this table, and here onward in the paper,
the provided error bars are corresponding to 1$\sigma$. Additionally, \citet{2026Gosset} also identified spectral signatures of a third component, termed as a potential ``intruder'' star. They estimated this intruder to have a spectral type between O6 and O8, although no direct confirmation of this star and its association to the WN6ha + O5 binary could be established.

Constraining the binary component masses of WR~25, along with establishing the nature of the third component, will be a crucial step in a comprehensive analysis of the system. In this paper, we present interferometric observations of WR~25 taken with the Very Large Telescope Interferometer (VLTI) at the European Southern Observatory (ESO) Chile. We spatially resolve the two binary components and report the first dynamical mass measurement of this system. We also detect the third component, likely gravitationally bound to the binary. Taking advantage of the brightness ratios derived from interferometry, we revisit the spectroscopic data of WR~25 and perform spectral disentangling to concomitantly separate the three components and compute their stellar parameters.
The paper is structured as follows:
Section\,\ref{sec:obs} provides a summary of the observations and data reduction. In Section\,\ref{sec:analysis}, we describe the analysis of interferometric data. Section\,\ref{sec:spedis} details our spectral disentangling approach, followed by computation of stellar parameters and orbital properties described in Section\,\ref{sec:CMFGEN}. Finally, we conclude in Section\,\ref{sec:disc} by discussing the nature of WR~25 and future prospects for studying the system.

\section{Observations and data reduction} 
\label{sec:obs}

Long-baseline interferometric observations for WR~25 were obtained in the infrared $H$ band (1.52 -- 1.76 $\mu\mathrm{m}$) with the VLTI/PIONIER instrument \citep{2011LeBouquin} using the four 1.8-m Auxiliary Telescopes (ATs) at ESO. The AT configuration used was A0-G1-J2-K0, offering a maximum baseline of about 130 m. WR~25 was observed on the night of January 16, 2019 (Heliocentric Julian Date or HJD 2,458,500.733). The reduced data products were retrieved from the ESO Data Archive (ESO Program ID 0102.D-0234(A), PI. Sana).

For spectral disentangling (Sect.\,\ref{sec:spedis}), we adopt the ``Data set I'' spectroscopic data from \citet{2026Gosset}. The data were taken using the FEROS spectrograph \citep{1999Kaufer} at the MPG/ESO 2.2-m telescope at La Silla, Chile. They span a broad wavelength range from 3700 \AA\ to 9200 \AA, with a spectral resolving power of $\approx$48,000. A detailed description of the data can be found in section 2.1 of \citet{2026Gosset}.

We also include an archival ultraviolet (UV) spectrum of WR~25 obtained with the \textit{International Ultraviolet Explorer} (\textit{IUE}) at Heliocentric Julian Date 2,444,139.591 (\textit{IUE} Program ID IGBJH, PI. Hesser). The UV spectrum covers the 1150 -- 1800 \AA\ wavelength range, and is particularly useful in determining stellar parameters of the primary.

\begin{table}
    \begin{center}
    \caption{Orbital parameters for WR~25 from spectroscopy.}
    \label{tab:orb}
    \begin{tabular}{ccc}
        \hline
        \hline
       Parameter & Value \\
       \hline
       $P$ [d] &  $207.638\pm0.093$ \\
       $e$ &  $0.595\pm0.013$ \\
       $\omega$ [deg] &  $212.04\pm1.59$ \\
       $T_0$ [d]* & $51184.08\pm0.46$ \\
       $K_1$ [km\,s$^{-1}$] &  $53.18\pm0.82$ \\
       $q$ [$M_1/M_2$] &  $2.02\pm0.36$ \\
       $f$ [M$_\odot$] &  $1.687\pm0.098$ \\
       \hline
    \end{tabular}
    \end{center}

\smallskip
\textbf{Notes:} Parameters include orbital period ($P$), eccentricity ($e$), argument of periastron ($\omega$), time of periastron passage ($T_0$) and primary radial velocity amplitude ($K_1$), primary-to-secondary mass ratio ($q$) and binary mass function ($f$). *: Heliocentric Julian Date 2,400,000+.
\end{table}

\section{Interferometric analysis} \label{sec:analysis}

The PIONIER interferometric data consist of two observables: squared visibilities (V2) and closure phases (T3PHI). V2 encodes information about how spatially resolved the source is and the flux contrast between different components. T3PHI, on the other hand, is sensitive to point-asymmetry within the source. An unresolved source yields V2 = 1 and T3PHI = 0 for all baselines and closure triangles respectively, while a spatially resolved source shows significant deviation in comparison. For binaries or higher order multiples, the PIONIER instrument is sensitive to separations of 1 -- 45 milliarcseconds (mas) and $H$-band flux contrasts down to $\Delta H\sim4$ \citep{2011Absil,2014Sana}.

\begin{figure}
    \centering
    \includegraphics[trim={23cm 1cm 0cm 1.8cm}, clip, width=0.95\linewidth]{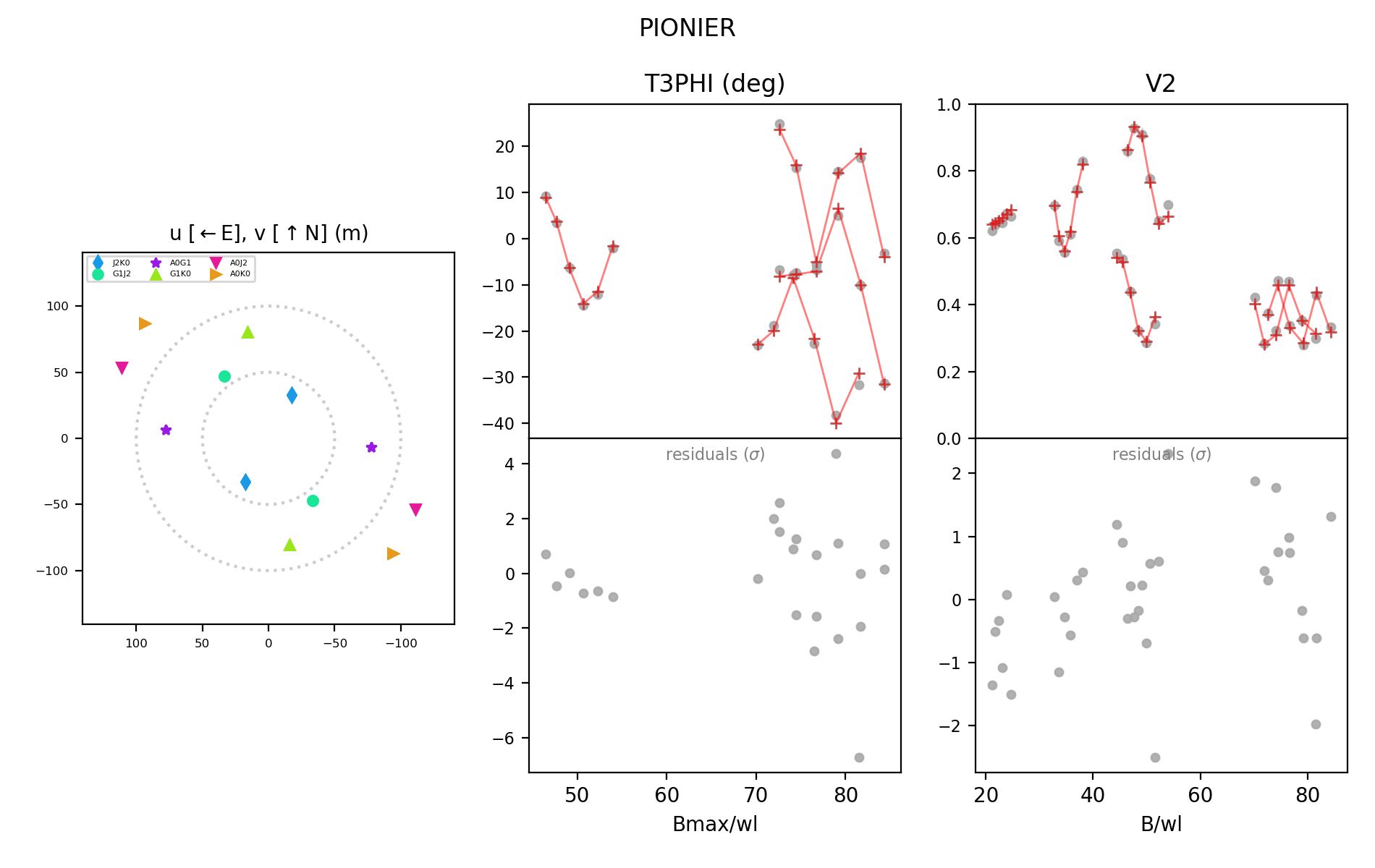}
    \caption{The closure phases (T3PHI) and squared visibilities (V2) shown for WR~25 as functions of spatial frequencies (baseline B in m/wavelength wl in $\mu$m). The data are shown in the upper panels with grey points while the best-fit triple component model is shown with red crosses. The points connected by red lines represent different 2-telescope baselines for V2 and 3-telescope triangles for T3PHI. The lower panels show residuals corresponding to the best-fit model.}
    \label{fig:triple}
\end{figure}

We use the interferometric modeling software PMOIRED\footnote{\url{https://github.com/amerand/PMOIRED}} \citep{2022Merand} version 1.2.10 for geometric modeling of WR~25 data to best explain the features seen in V2-T3PHI.  On visual inspection, it is clear that V2 and T3PHI deviate significantly from the signature of an unresolved source, hinting at the target being spatially resolved (see Fig.\,\ref{fig:triple}). However, as a baseline model, we first try to fit the data with a single uniform disk component with its diameter allowed to vary. The best fit was poor as expected, resulting in a reduced chi-squared ($\chi^2_{\rm red}$) of 747.

We then add a level of complexity and try a binary star model. This is implemented by fixing the position of one unresolved component at the origin, and allowing the position of a second unresolved component to vary over a grid of points around the central star covering the full PIONIER field of view \citep{2015Gallenne}. Both components were modeled with uniform disks of fixed diameter 0.2 mas (thus unresolved). The fluxes of the two components were allowed to vary, with their sum fixed to 1. We used a square grid of points ranging from $-45$ to 45 mas in both N and E directions in steps of 1 mas. The best fit resulted in $\chi^2_{\rm red}$ = 234, which was better than the single component fit but still unsatisfactory.

\begin{figure}
    \centering
    \includegraphics[trim={0cm 0cm 3cm 1cm}, clip, width=0.95\linewidth]{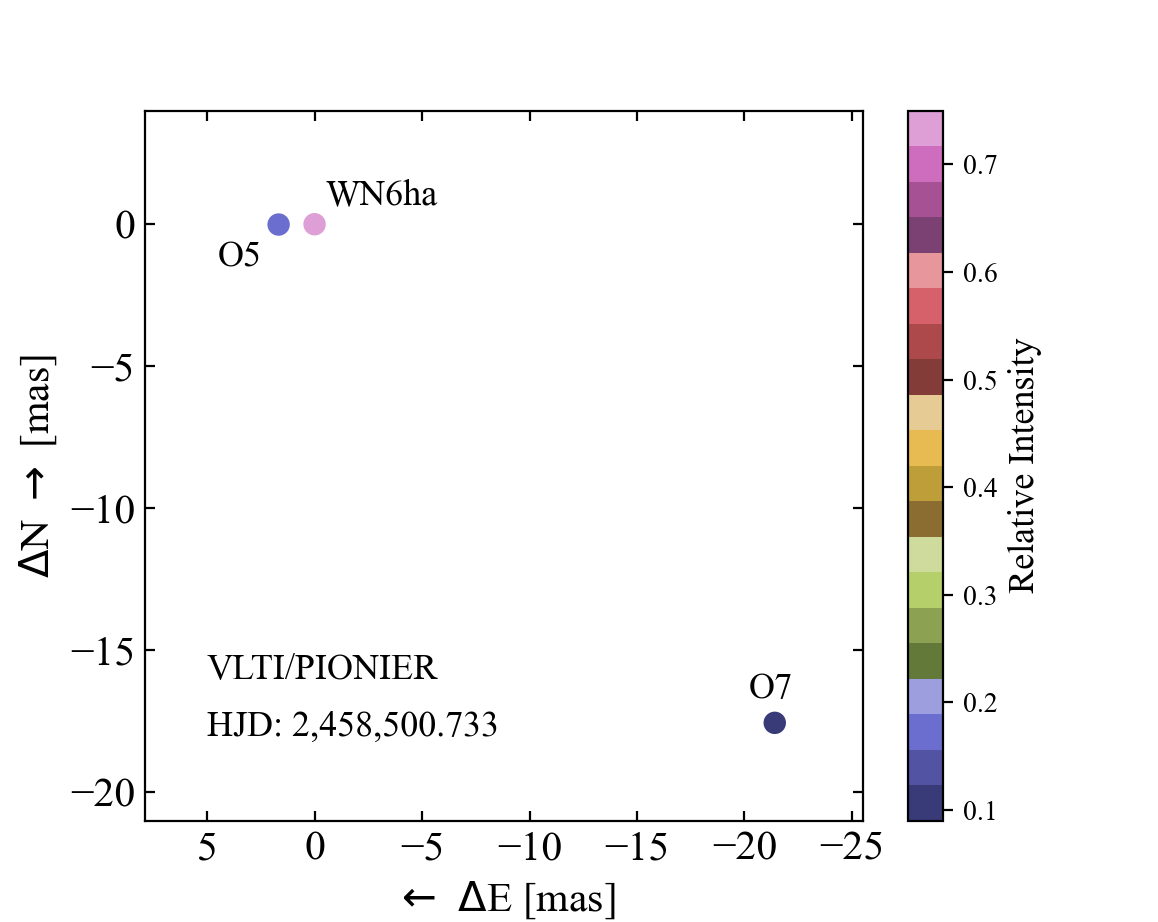}
    \caption{Relative positions of the three components in WR~25. All error bars are smaller than the marker sizes and the colors indicate relative intensities of the components measured in the $H$-band by PIONIER, as listed in Table\,\ref{tab:triple}. Component spectral types derived from spectroscopic analysis are also shown (more details in Sect.\,\ref{sec:spedis}).}
    \label{fig:pos}
\end{figure}

To test the possible presence of a third star, we add one more component to try a triple component model. This was done similarly to the binary model, this time with the position of one component at the origin while the other two varied over their own grids. With the prior knowledge that WR~25 consists of a 208-day WN6ha + O5 binary, one of the two companions could potentially be the O5 companion, expected to be within a few mas of the primary \citep{2026Gosset}. Consequently, we set a small grid ranging from $-3$ to 3 mas with steps of 0.3 mas in both N and E directions to search for the O5 companion. For the third component, we set a coarser, more agnostic grid from $-45$ to 45 mas with steps of 1 mas in both N and E directions. 
The triple component best-fit is excellent as shown in Figure\,\ref{fig:triple}, explaining both V2 and T3PHI well, with $\chi^2_{\rm red}$ = 2.84. The final model consists of (i) the central star as the brightest component, the luminous WN6ha star; (ii) a close component at $1.68\pm0.02$ mas, the spectroscopic O5 companion; and (iii) a farther out component at $27.69\pm0.02$ mas, likely an outer companion consistent with the intruder reported by \citet{2026Gosset}. Both companions are detected at a significance greater than $8\sigma$ with respect to the baseline single component model. To obtain realistic error bars for all the best-fit parameters, we used a bootstrapping routine with 1000 realizations. The results are listed in Table\,\ref{tab:triple} along with component spectral types (see Sect.\,\ref{sec:spedis} for details). A visual representation on the plane of the sky of the three components is shown in Figure\,\ref{fig:pos}.

While the secondary is a known companion from spectroscopy, we present the first unambiguous detection of the third component using interferometry. 
Although one snapshot observation is not sufficient to determine whether the star is gravitationally bound to the inner binary, we can assess its probability of chance alignment. Following \citet{2020Rainot}, we calculate the probability of this component being spuriously associated to the inner binary ($P_{\rm spur}$). To do so, we query all sources from the 2MASS All-Sky Point Source Catalog \citep{2003Skrutskie} within 120" and with $\Delta H < 5$ (one magnitude fainter than typical PIONIER sensitivity) with respect to WR~25 to conservatively estimate the density of nearby sources bright enough for detection with PIONIER. We then calculate the probability $P_{\rm spur}$ of spuriously detecting a source within the PIONIER outer working angle \citep[$\sim$45 mas,][]{2014Sana}. We find $P_{\rm spur}<0.01\%$ for the third component, implying a very high likelihood of it being physically related to the inner binary. In Table\,\ref{tab:triple} and in the following text, we systematically name this object the tertiary component despite the persisting possible doubt.

\begin{table*}[]
    \begin{center}
    \caption{Astrometric and photometric constraints on WR~25.}
    \label{tab:triple}
    \renewcommand{\arraystretch}{1.15}
    \begin{tabular}{cccccccc}
        \hline
        \hline
       Component & $\Delta E$ [mas] & $\Delta N$ [mas] & $e_{\rm maj}$ [mas] & $e_{\rm min}$ [mas] & PA [deg] & $f$ & SpT\\
       \hline
       Primary & 0* & 0* & - & - & - & $0.751\pm0.003$ & WN6ha \\
       Secondary & $1.682$ & $-0.014$ & 0.020 & 0.007 & $-68.4$ & $0.158\pm0.002$ & O5 \\
       Tertiary & $-21.410$ & $-17.557$ & 0.025 & 0.011 & $-34.1$ & $0.091\pm0.002$ & O7 \\
       \hline
    \end{tabular}
    \end{center}

\smallskip
\textbf{Notes:} Relative astrometric positions ($\Delta E$ and $\Delta N$ in mas) and relative flux contributions ($f$, dimensionless) of all three components in WR~25. Marked by *, the position of the primary (WR) is fixed at the origin. For the secondary and tertiary, the error ellipse is provided in the form of the semi-major axis ($e_{\rm maj}$), semi-minor axis ($e_{\rm min}$), and position angle (PA) of the semi-major axis measured eastward of north. Also listed are component spectral types (SpT), which are discussed in Section\,\ref{sec:spedis}.
\end{table*}

\section{Spectral disentangling}
\label{sec:spedis}

With the newly confirmed presence of the third component, we turn to spectroscopic data of WR~25 taken with FEROS \citep{2026Gosset} to isolate the spectral signatures of individual components. We employ a three-component Fourier spectral disentangling code adapted from \citet{Ilijic2004} and \citet{Hadrava1995}. For the inner system, we adopt the orbital solution from Table\,\ref{tab:orb}, and assume that the third component remains stationary \citep[see][for details]{2026Gosset}. The flux contributions of the individual components are taken from interferometric measurements, with 
$f_\mathrm{P} = 0.751$, $f_\mathrm{S} = 0.158$, and $f_\mathrm{T} = 0.091$ for the primary, secondary, and tertiary, respectively (Table\,\ref{tab:triple}). 
Given the O-star natures of all three components, it is safe to assume that they are in the Rayleigh-Jeans domain from the UV to the infrared. In principle, free-free excess stemming in the wind of the WR star could cause significant infrared excess in the $H$-band. However, this has negligible impact in the case of WR~25 due to the primary's relatively weak wind, as we show below from a detailed spectral analysis of the individual components of the system (Sect.~\ref{sec:CMFGEN}). Hence, we assume that the $H$-band flux ratios represent the ratios in the $V$-band and UV as well.

Because the spectral lines of the secondary O5 star never fully deblend from those of the WN6ha primary, particularly when blue-shifted, the reconstruction of its spectral lines can exhibit artifacts in their blue wings. In addition, the spectral lines of the tertiary component are mostly overlapping with those of the primary, with only a few exceptions, such as the \ion{He}{i}~4471 line, where the almost static signature of the tertiary has been identified in the observed spectra \citep{2026Gosset}. The disentangled spectrum of the tertiary therefore shows only clear spectroscopic features of a limited number of lines, including \ion{He}{i}~4471 and \ion{He}{ii}~4542. Given the broadness of the Balmer lines, and the fact that these lines never fully deblend for the three components, only a faint contribution can be extracted for the tertiary from the disentangling process. While \ion{He}{i}~4471 and \ion{He}{ii}~4542 are clearly resolved in the disentangled spectrum, the shape of the Balmer lines could only be reconstructed using the continuum of synthetic spectra with the same spectral classification as the tertiary. Consequently, the determination of the surface gravity and wind properties for the tertiary remains highly uncertain (see Sect.~\ref{sec:CMFGEN}).

As noted by \citet{Mahy2017}, Fourier spectral disentangling suffers from the drawback of losing the continuum for stars that are not fully eclipsed by their companions. It is therefore crucial to correct the disentangled spectra for this effect. For the WN6ha primary, the continuum is particularly difficult to constrain, so we corrected for continuum variations by fitting a spline to a carefully-selected continuum region using synthetic models. For the secondary star, the continuum correction was also performed by fitting a spline through its continuum. For the tertiary component, as mentioned above, only the regions around the Balmer lines were defined using synthetic spectra; for the remaining portions, the continuum was fitted using a spline function, as was done for the secondary.

The WN6ha spectrum (see Fig.\,\ref{fig:WN_CMFGEN}) is characterized by a prominent \ion{N}{iv}~4058 line and a strong \ion{N}{v} doublet at 4604 -- 4619~\AA. The \ion{He}{i}~4471 line is not detected in the disentangled spectrum, while the \ion{Si}{iv}~4089 -- 4116 lines are observed in emission. The \ion{He}{ii}~4686 line is also strongly in emission.

In the spectrum of the O5 secondary (see Fig.\,\ref{fig:CMFGEN_secondary}), the \ion{Si}{iv}~4089 line is absent, while \ion{Si}{iv}~4116 is clearly detected in emission. The \ion{N}{iii}~4634–4641 doublet appears in strong emission, and \ion{He}{ii}~4686 is observed slightly in absorption. These characteristics suggest a classification with the suffix ((f$^{+}$)) or (f$^{+}$) \citep[][see also the discussion in section 7.2
of \citealt{2026Gosset}]{Walborn1972}.

The disentangled spectrum of the tertiary component (see Fig.\,\ref{fig:CMFGEN_tertiary}) displays clear \ion{He}{i} and \ion{He}{ii} features, indicating a mid-O classification. The \ion{He}{i}~4471 line has a depth comparable to that of \ion{He}{ii}~4542, suggesting a spectral type of O7–O7.5. Since \ion{He}{ii}~4686 is strongly in absorption, the tertiary can be assigned a luminosity class V.
We refer to this component as an O7 star here onward.

\section{Stellar parameters and evolutionary status}
\label{sec:CMFGEN}

\subsection{Atmosphere modeling}

The spectra of the three components were modeled using the non-local thermodynamic equilibrium (non-LTE) radiative transfer code CMFGEN \citep{Hillier1998}. CMFGEN solves the radiative transfer equation in the co-moving frame under the assumptions of spherical symmetry and steady state, accounting for line blanketing and complex atomic models in expanding stellar winds. The code simultaneously solves the equations of statistical equilibrium and radiative transfer, allowing for a detailed treatment of millions of spectral lines through a super-level approach to reduce computational complexity. Wind structures are prescribed via a $\beta$-type velocity law, and clumping can be incorporated through a volume-filling factor with a void inter-clump medium formalism. The temperature structure is obtained from radiative equilibrium, ensuring consistency between the radiation field and the atmospheric stratification.

\subsubsection{The WN6ha primary}

To model the WN6ha primary spectrum, we generated, as a first step, a grid of about 1000 models covering the parameter space: from 42~kK to 50~kK with steps of 2~kK for the effective temperature, from 0.10 to 0.30, with steps of 0.04 for the helium surface abundance, from 8.0 to 10.0 with steps of 0.5 for the nitrogen surface abundance (in $\epsilon_{\rm N}= 12 + \log [\mathrm{N/H}]$ notation), and from --5 to --4, with steps of 0.2 for the mass-loss rate in logarithmic scale ($\log \dot{M}$ [$M_{\odot}\,\mathrm{yr}^{-1}$]). Other parameters were kept fixed at first: clumping volume filling factor of 0.15, $\beta$ exponent of the velocity law of 1.0, terminal wind speed $v_{\infty} = 2500$~km~s$^{-1}$, and stellar luminosity of $\log(L/L_{\odot}) = 6.0 $ in agreement with \citet{2026Gosset}. We note that we do not perform a Spectral Energy Distribution fitting given the incompatibility between the extinction values provided from the blue/UV, and red/infrared \citep[see][]{crowther95,2026Gosset}. Using this grid, we computed the $\chi^2$ statistic to compare the disentangled spectrum of the primary with the different models and narrow down the parameter space around a set of representative models.

\begin{figure}
    \centering
    \includegraphics[width=0.95\linewidth]{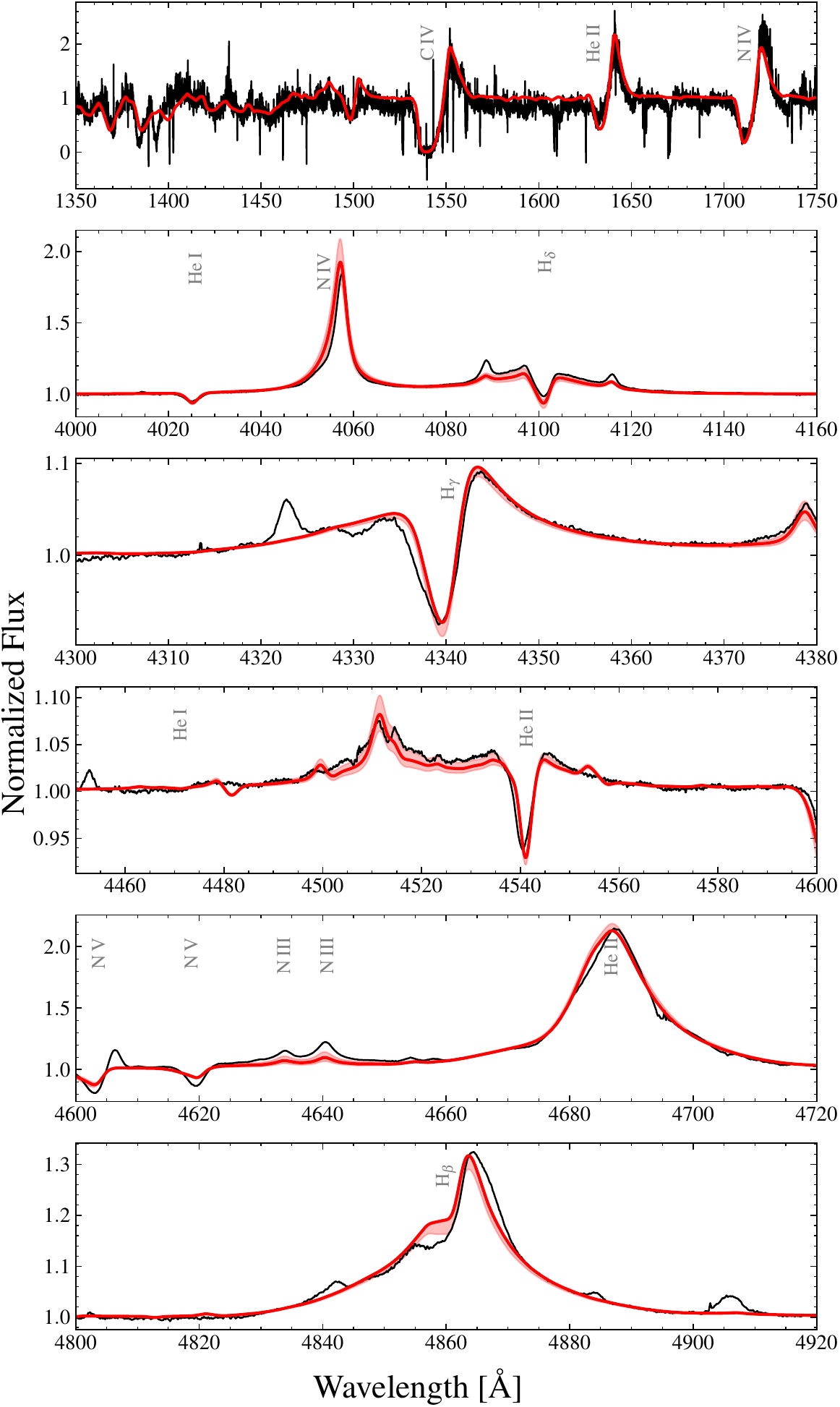}
    \caption{Comparison between the best-fit CMFGEN model (in red, and the uncertainty region
    in light-red) with either the observed UV spectrum of WR~25 (in black, top panel) 
    or with the disentangled spectrum of the primary component of WR~25 (in black, 
    the five lower panels).}
    \label{fig:WN_CMFGEN}
\end{figure}

In a second step, we refined the stellar and wind parameters through an iterative fitting procedure aimed at converging toward the optimal solution. Starting from the representative models identified from the $\chi^2$ analysis, we progressively adjusted the effective temperature, surface abundances, and wind properties to improve the agreement between the synthetic and observed spectra. To alleviate degeneracies between key parameters, we complemented the disentangled spectrum with the ultraviolet composite spectrum of WR~25 obtained with \textit{IUE}, that includes lines like \ion{N}{v}~1240, \ion{O}{v}~1371, the \ion{Si}{iv}~1393–1402 doublet, \ion{C}{iv}~1548–1550, \ion{He}{ii}~1640, and \ion{N}{iv}~1718. By adding the UV lines, we add more diagnostic lines to constrain the mass-loss rate, and we can now derive the terminal wind velocity for the primary by using the blue edges of the P-Cygni profiles (i.e., the \ion{C}{iv}~1548–1550, \ion{He}{ii}~1640, and \ion{N}{iv}~1718 lines). For this purpose, we shifted the synthetic spectra of the WN6ha primary and the O5 secondary according to their radial velocities calculated from the orbital solution presented by \citet{2026Gosset}. For the tertiary, we use the same radial velocity as the primary star. We also correct for the dilution effects in the UV (see Sect.\,\ref{sec:spedis}). The best-fit model is shown in 
Fig.\,\ref{fig:WN_CMFGEN} and is compared to the composite \textit{IUE} spectrum in the UV and to the disentangled spectrum in the optical. The temperature of the WN6ha primary was determined using the nitrogen ionization balance (\ion{N}{iii}/\ion{N}{iv}/\ion{N}{v}). However, the \ion{N}{v} lines at 4604 and 4619~\AA\ proved difficult to reproduce while simultaneously fitting the \ion{N}{iii} lines present in the disentangled spectrum (equal weight was given to both sets of ionization stages). The best-fit temperature (at $\tau = 20$) for the WN6ha star is $43 \pm 1$~kK, providing an effective temperature of $42.2 \pm 1.0$~kK (at $\tau = 2/3$). We derived a helium surface abundance of $Y_{\rm He} = 0.14 \pm 0.02$, a nitrogen abundance of $\epsilon_{\rm N} = 8.90 \pm 0.08$, and carbon and oxygen abundances of $\epsilon_{\rm C} = 6.78 \pm 0.12$ and $\epsilon_{\rm O} = 8.18 \pm 0.43$, respectively. The mass-loss rate is $\log \dot{M}\,[M_{\odot}\,\mathrm{yr}^{-1}] = -5.15 \pm 0.10$, and the terminal wind velocity is derived to be $v_{\infty} = 2200 \pm 100$~km\,s$^{-1}$, from the blue edges of the UV P-Cygni profiles, such as \ion{C}{iv}~1548 -- 1550, \ion{He}{ii}~1640, and \ion{N}{iv}~1718. The clumping volume-filling factor is $\approx0.1$, and the exponent of the velocity law is $\beta \approx 1.5$. We also computed the bolometric correction using the calibration of \citet{2006Martins} and adopted the absolute $H$-band magnitudes of the individual components. This yields a luminosity of $\log(L/L_{\odot}) = 6.21 \pm 0.06$. All best-fit parameters are listed in Table\,\ref{tab:parameters}. 

\subsubsection{The O5 secondary and O7 tertiary}

We first derived the projected rotational velocities of the secondary and tertiary components. For this purpose, we used a Python implementation of the IACOB-broad tool developed by \citet{Simon-Diaz2007,Simon-Diaz2014}, which we adapted for our analysis. For the O5 secondary, we obtained a projected rotational velocity of $v\sin i = 87 \pm 12$~km\,s$^{-1}$ and a macroturbulent velocity of $v_{\rm mac} = 69 \pm 34$~km\,s$^{-1}$. For the tertiary component, we derived $v\sin i = 69 \pm 24$~km\,s$^{-1}$ and $v_{\rm mac} = 83 \pm 52$~km\,s$^{-1}$.

\begin{figure*}
    \centering
    \includegraphics[width=0.5\linewidth]{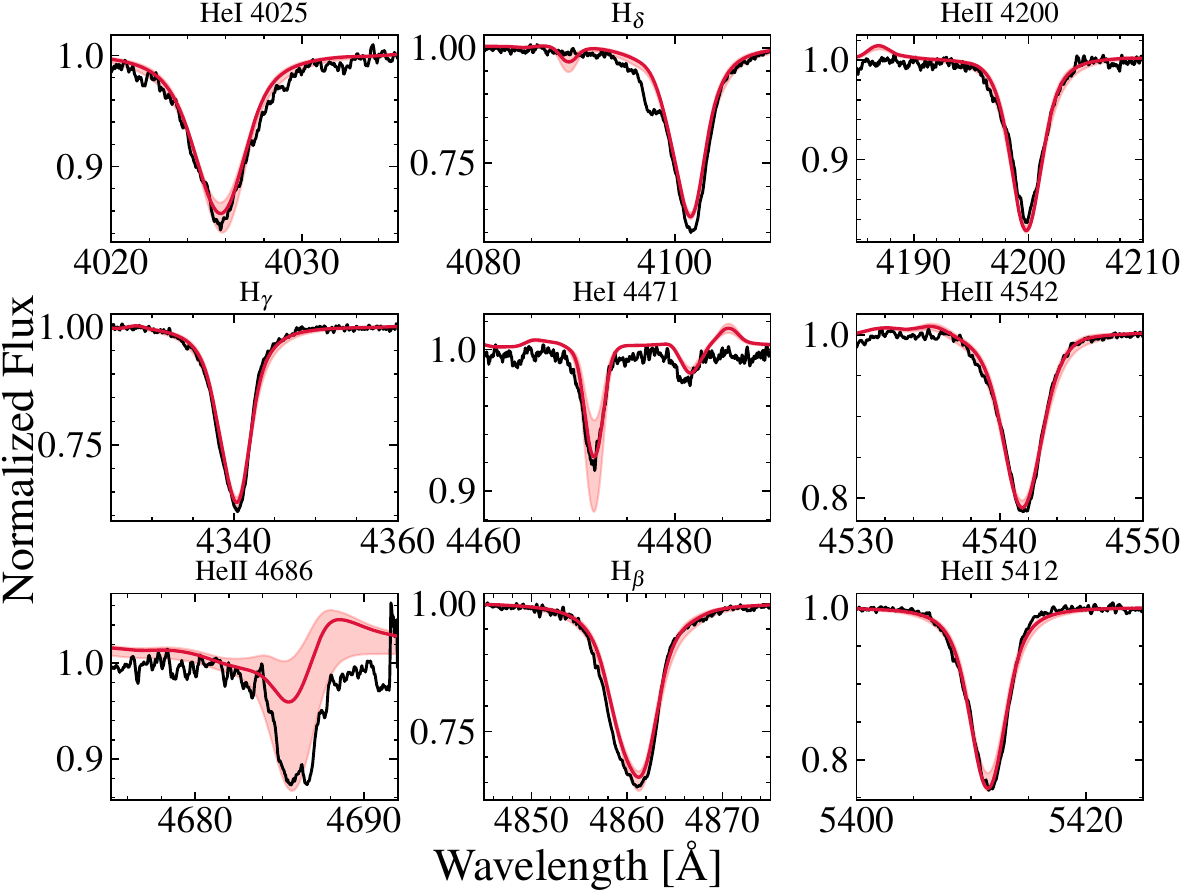}
    \hfil
    \includegraphics[width=0.495\linewidth]{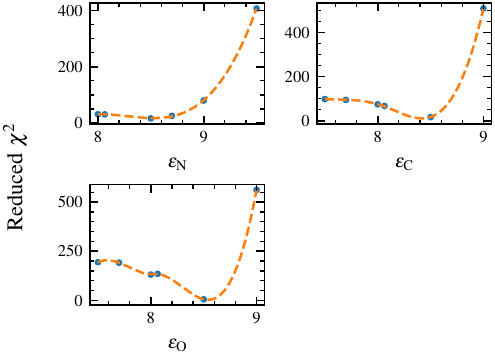}
    \caption{Left: Comparison between the disentangled spectrum (black) of the secondary component of WR~25 with the best-fit CMFGEN model (solid red) and the uncertainty region (shaded red), shown for nine prominent H and He lines. Right: Reduced $\chi^2$ plots for nitrogen, carbon, and oxygen surface abundances of the secondary expressed in $\epsilon_{\rm X}= 12 + \log [\mathrm{X/H}]$ formalism.}
    \label{fig:CMFGEN_secondary}
\end{figure*}

To estimate the fundamental parameters of the secondary and tertiary components, we used a pre-computed grid of synthetic spectra calculated with CMFGEN. The grid was constructed based on the evolutionary tracks of \citet{2011Brott}. We adopted a constant step of 0.1~dex in surface gravity ($\log g$), while adjusting the effective temperature, luminosity, and mass-loss rate consistently with the evolutionary models. The terminal wind velocity was estimated using the $T_{\rm eff}-v_\infty$ prescription of \citet{hawcroft2024}. Surface abundances were fixed to solar values following \citet{Grevesse2010}.

All models were computed with different microturbulent velocities of 2, 5, 7, 10, 12, 15, 17, and 20~km\,s$^{-1}$. The clumping volume filling factor was fixed to 0.1, while the exponent of the velocity law $\beta$ was fixed to 1.0. The grid contains 5271 models, covering $T_{\rm eff}$ between 10 and 50~kK, surface gravities in the range $\log g = 1.4$–4.3 (cgs), and luminosities between $\log(L/L_{\odot}) = 2.60$ and 5.92.

\begin{figure*}
    \centering
    \includegraphics[width=0.5\linewidth]{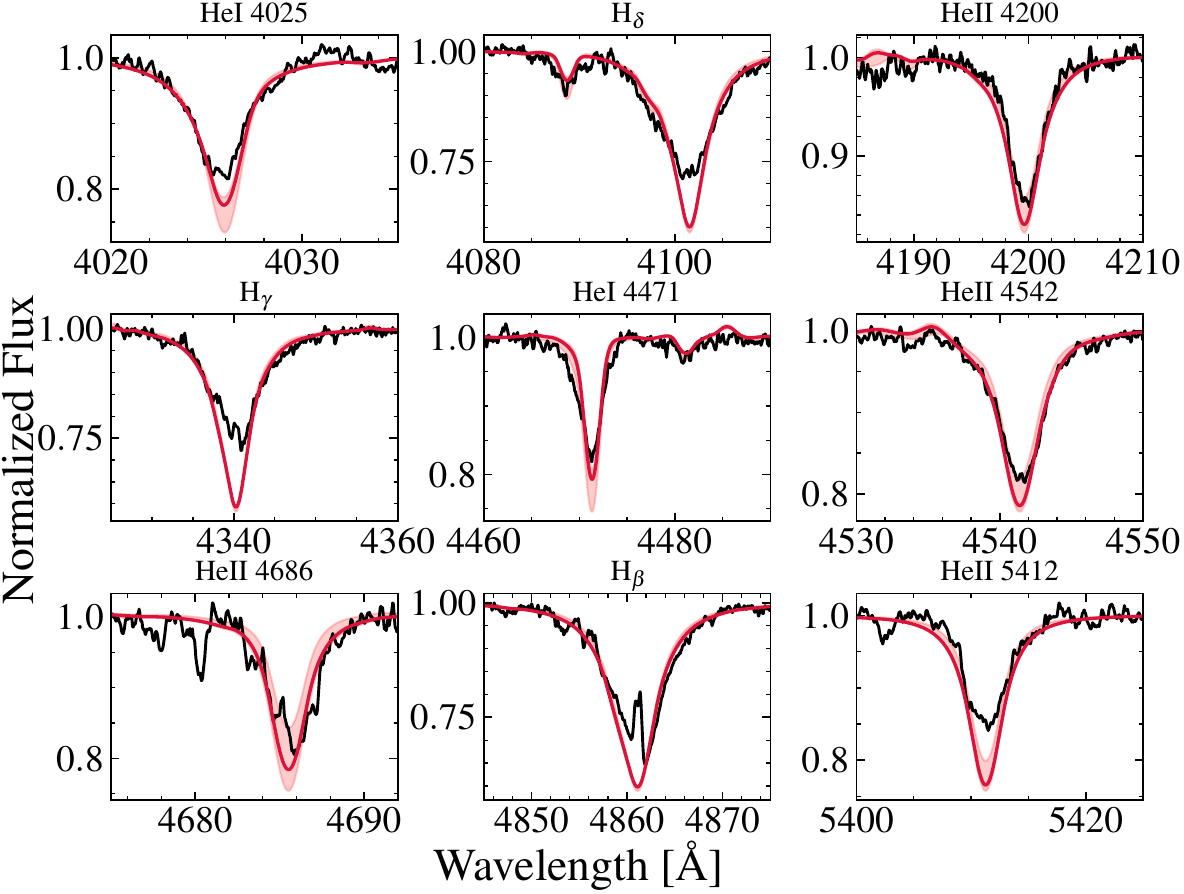}
    \hfil
    \includegraphics[width=0.48\linewidth]{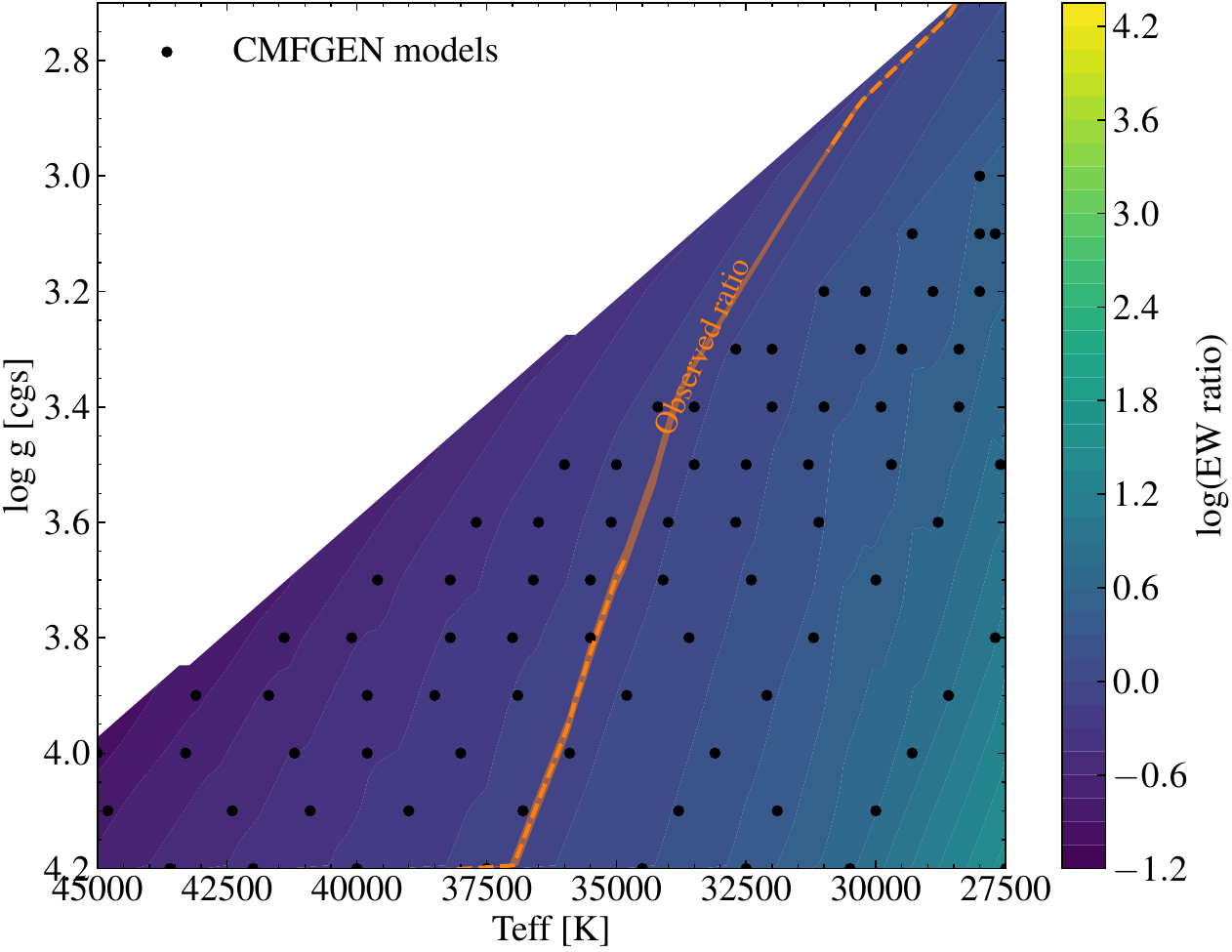}
    \caption{Left: Same as for the left set of panels of Fig.~\ref{fig:CMFGEN_secondary} but for the tertiary component of WR~25. Right: Kiel diagram ($\log g$ vs $T_{\rm eff}$) computed from the CMFGEN grid for the tertiary component of WR~25. Each point represents a synthetic model included in the grid. The orange region indicates the models for which the predicted equivalent-width ratio of \ion{He}{i}~4471 to \ion{He}{ii}~4542 matches the observed ratio within uncertainties.}
    \label{fig:CMFGEN_tertiary}
\end{figure*}

The effective temperature was determined from the helium ionization balance, while the surface gravity was constrained from the wings of the Balmer lines, in particular H$\delta$, H$\gamma$, and H$\beta$. The best-fitting model was selected by minimizing the $\chi^2$ function computed over the hydrogen and helium lines. The left panels of Figures~\ref{fig:CMFGEN_secondary} and \ref{fig:CMFGEN_tertiary} present the best-fit grid models for the secondary and tertiary components, together with their associated uncertainty regions.

Given the uncertainty in the continuum correction around the Balmer-line wings in the disentangled tertiary spectrum, we adopted an additional diagnostic to estimate the stellar parameters of this component. We measured the equivalent widths of the \ion{He}{i}~4471 and \ion{He}{ii}~4542 lines and computed their ratio. This observed ratio was then compared to the ratios predicted by all synthetic spectra in our CMFGEN grid. Models for which the theoretical and observed ratios agree within uncertainties were retained as viable solutions. The right panel of Fig.~\ref{fig:CMFGEN_tertiary} shows the Kiel diagram constructed from our grid models, with the region consistent with the observed line ratio highlighted in orange. This constraint is independent of the brightness-ratio correction applied to the disentangled tertiary spectrum. The acceptable parameter space for the tertiary, assuming the same distance and coevality with the binary, is located in the range $34 \leq T_{\rm eff} \leq 38$~kK and $3.4 \leq \log g \leq 4.2$. The luminosities of the O5 secondary and O7 tertiary were estimated from the individual absolute magnitudes corrected from the bolometric correction given by \citet{2006Martins} using the effective temperatures of the stars and the $H$-band law.

Once the best-fitting models were established, we derived the surface abundances by computing additional spectra with varying carbon, nitrogen, and oxygen abundances. For that purpose, we use the \ion{N}{iii}~4097, the \ion{N}{iii}~4198, the \ion{N}{iii} quadruplet around 4515~\AA, the \ion{C}{iii}~4068$-$70, and the \ion{O}{iii}~5592 lines. The optimal values were determined by minimizing the $\chi^2$ function computed on these selected diagnostic lines and is shown for the secondary, for example, in the right panel of Fig.~\ref{fig:CMFGEN_secondary}. The O5 secondary exhibits nitrogen enrichment, while its carbon and oxygen abundances are consistent with solar values. The O7 tertiary shows surface abundances compatible with solar values for carbon, nitrogen, and oxygen. All the individual parameters for the three components are listed in Table~\ref{tab:parameters}.

\begin{table}
\centering
\caption{Individual parameters of the WR~25 components and their $1\sigma$ uncertainties.}
\label{tab:parameters}
\begin{tabular}{lccc}
\hline
\hline
 & Primary & Secondary & Tertiary \\
\hline
\multicolumn{4}{c}{Stellar parameters} \\
\hline
Sp.\ Type & WN6ha & O5\,((f$^{+}$)) & O7\,((f)) \\
$T_{\rm eff}$ [kK] 
    & $42.2 \pm 1.0$ 
    & $41.4 \pm 1.7$ 
    & $36.3 \pm 2.6$ \\
$T_{\star}$ [kK] 
    & $43.0 \pm 1.0$ 
    & $42.2 \pm 1.7$ 
    & $36.8 \pm 2.6$ \\
$\log g$ [cgs] 
    & -- 
    & $3.9 \pm 0.1$ 
    & $3.9^{+0.1}_{-0.2}$ \\
$\log (L/L_{\odot})$ 
    & $6.21 \pm 0.06$ 
    & $5.48 \pm 0.07$ 
    & $5.08 \pm 0.10$ \\
$\xi_t$ [km\,s$^{-1}$] 
    & $20 \pm 5$ 
    & $15 \pm 3$ 
    & $10 \pm 2$ \\
\hline
\multicolumn{4}{c}{Wind parameters} \\
\hline \\[-10pt] 
$\log \dot{M}$ [$M_{\odot}\,\mathrm{yr}^{-1}$] 
    & $-5.15 \pm 0.10$ 
    & -- 
    & -- \\
$v_{\infty}$ [km\,s$^{-1}$] 
    & $2200 \pm 100$ 
    & -- 
    & -- \\
$\beta$ 
    & 1.5 
    & 1.0 
    & 1.0 \\
$f_{\rm cl}$ 
    & 0.10 
    & 0.10 
    & 0.10 \\
\hline
\multicolumn{4}{c}{Surface abundances} \\
\hline
$Y_{\rm He}$ 
    & $0.14 \pm 0.02$ 
    & $0.09 \pm 0.01$ 
    & $0.09 \pm 0.01$  \\
$\epsilon_{\rm N}$ 
    & $8.90 \pm 0.08$ 
    & $8.65 \pm 0.08$  
    & $7.65 \pm 0.10$  \\
$\epsilon_{\rm C}$ 
    & $6.78 \pm 0.32$ 
    & $8.45 \pm 0.08$  
    & $8.39 \pm 0.14$  \\
$\epsilon_{\rm O}$ 
    & $8.18 \pm 0.43$ 
    & $8.58 \pm 0.08$  
    & $8.59 \pm 0.23$  \\
\hline
\end{tabular}

\end{table}

\subsection{Dynamical mass constraints}
\label{sec:masstot}

In a spatially resolved binary, relative positions of the two components measured at different orbital phases can be combined with a known distance and spectroscopic orbital solution of one of the two components to obtain a complete 3D orbit and the total mass of the system \citep[e.g.][]{2024Frost}. In cases where the mass ratio is known, individual component masses can also be derived. Alternatively, a double-lined spectroscopic (SB2) orbital solution including an inclination estimate can be combined with a single interferometric epoch for a direct measurement of the dynamical masses \citep[e.g.][]{2025Deshmukh}. The data available for WR~25 fall short of either condition, yet the fortunate timing of the VLTI observation still allows us to constrain the total mass of the system without invoking the poorly constrained RV solution of the secondary.

Adopting the ephemeris of the primary RV solution given in Table~\ref{tab:orb},  the interferometric epoch of observation corresponds to an orbital phase of $\phi_\mathrm{int}=0.235\pm0.016$, where an orbital period uncertainty of 0.093~d from \citet{2026Gosset} has been included in the error propagation. This yields a true anomaly  of $\nu_\mathrm{int}=144.6\pm3.1\degr$ and $\nu_\mathrm{int}+\omega=356.6\pm3.5\degr$. The VLTI observation therefore occurred close to the passage at one of the nodes (given by $\nu+\omega=k\pi$).  The positive primary RV at $\phi_\mathrm{int}\approx0.24$ informs us that $\nu+\omega=0$ actually corresponds to the ascending node of the primary.  
Interestingly, the secondary is almost directly East of the primary (Table~\ref{tab:triple}), i.e. with a position angle of $\approx 90\degr$. While not necessary for the total mass constraint, it is then straightforward to estimate the value of the longitude of the ascending node $\Omega$ to be $\approx 270$\degr. 

This favorable timing allows us to directly estimate the total mass of the system using only the primary RV curve and the distance, without requiring the mass-ratio that is plagued with significant uncertainty. At a node, for a distance $D$ to the system, the true separation $r$ and the projected physical separation $\rho_\mathrm{phys}=D \rho=D\sqrt{\Delta E^2+\Delta N^2}$ are identical to one another. This results from the projected separation equation 
\begin{equation}
    \rho_\mathrm{phys}^2=r^2 \left( \cos^2(\nu+\omega)+\sin^2(\nu+\omega) \cos^2 i \right),
\end{equation}
which simplifies to $\rho_\mathrm{phys}\approx r$ for $\nu+\omega=k\pi$ (with $k=0,1,2,\dots$). We thus obtain $\rho_{\rm phys}=3.90\pm0.21$ au.
This gives us an estimate of the absolute size of the relative orbit, without the need of a precise SB2 solution or additional VLTI measurements as 
\begin{equation}
    a = \rho_\mathrm{phys} \frac{1+e \cos \nu_\mathrm{int}}{1-e^2},
\end{equation}
where $\nu_\mathrm{int}$ is the true anomaly corresponding to the time of the interferometric observation, leading to $a=3.11\pm0.20$~au.
Constraints on the total mass then follow from the third Kelper law:
\begin{equation}
M_\mathrm{tot}=\frac{4\pi}{\mathrm{G}P^2}\left( \rho_\mathrm{phys}\frac{1+e \cos \nu_\mathrm{int}}{1-e^2} \right)^3,
\end{equation}
yielding a total mass of  $M_\mathrm{tot}=93\pm18$~\Msol. Combining with the mass ratio from Table\,\ref{tab:orb}, we get individual component masses $M_1=62\pm13\,M_\odot$ and $M_2=31\pm7\,M_\odot$.

We further leverage the node observation to constrain the inclination as, at the position of the node, we have 
\begin{equation}
    \sin i \approx \frac{a_1\sin i\cdot (1+q)}{\rho_\mathrm{phys}}\cdot\frac{1-e^2}{1+e\cos \nu_\mathrm{int}},
\end{equation}
where $q = 2.02\pm0.36$ (Table\,\ref{tab:orb}). This results in $\sin i = 0.79\pm0.11$, and $i \approx 52\pm10$\degr or $128\pm10$\degr. More interferometric observations will be necessary to break the degeneracy and enable a full 3D orbital solution.

\subsection{Evolutionary masses and ages}

To estimate the evolutionary masses and ages of the three components, we used the {\sc BONNSAI} (BONN Stellar Astrophysics Interface) Bayesian tool \citep{Schneider2014}. {\sc BONNSAI} compares the observed stellar properties with the BONN single-star evolutionary models \citep{2011Brott}.

In a first step, we performed a run using as input parameters the luminosity and effective temperature for the primary component, and the luminosity, surface gravity, effective temperature, and projected rotational velocity for the secondary and tertiary components. This yields evolutionary masses of $M_{\rm evol} = 87.2_{-11.0}^{+2.7}$, $39.0_{-2.6}^{+3.6}$, and $25.0_{-2.1}^{+2.6}\,M_\odot$ for the primary, secondary, and tertiary, respectively. The corresponding evolutionary ages are $1.2_{-0.2}^{+0.9}$, $1.9_{-0.5}^{+0.4}$, and $3.4_{-0.9}^{+0.9}$~Myr, respectively.

Given the large uncertainty on the surface gravity of the tertiary component, we performed an additional run excluding this parameter. In this case, we obtain an evolutionary mass of $25.6_{-2.3}^{+2.8},M_\odot$ and an age of $3.0_{-1.7}^{+1.2}$~Myr, consistent with the aforementioned values. We adopt these updated values for the tertiary, also used in Section\,\ref{sec:ter_orb}.

In a second step, we included the dynamical masses derived for the primary and secondary as additional constraints. {\sc BONNSAI} then returns evolutionary masses of $M_{\rm evol} = 75.2_{-4.7}^{+8.2}\,M_\odot$ and an age of $2.0_{-0.8}^{+0.2}$~Myr for the primary, and $M_{\rm evol} = 38.0_{-2.6}^{+3.1}\,M_\odot$ with an age of $2.0_{-0.5}^{+0.5}$~Myr for the secondary.

Finally, we included, in addition to the dynamical masses, the surface nitrogen abundances of the primary and secondary components as input parameters. In this configuration, {\sc BONNSAI} yields evolutionary masses of $M_{\rm evol} = 76.6_{-4.9}^{+3.1}\,M_\odot$ and an age of $2.0_{-0.2}^{+0.2}$~Myr for the primary, and $M_{\rm evol} = 37.0_{-1.8}^{+2.4},M_\odot$ with an age of $2.8_{-0.5}^{+0.5}$~Myr for the secondary.

\begin{figure}
    \centering
    \includegraphics[trim={0cm 0cm 0.5cm 0cm}, clip, width=\linewidth]{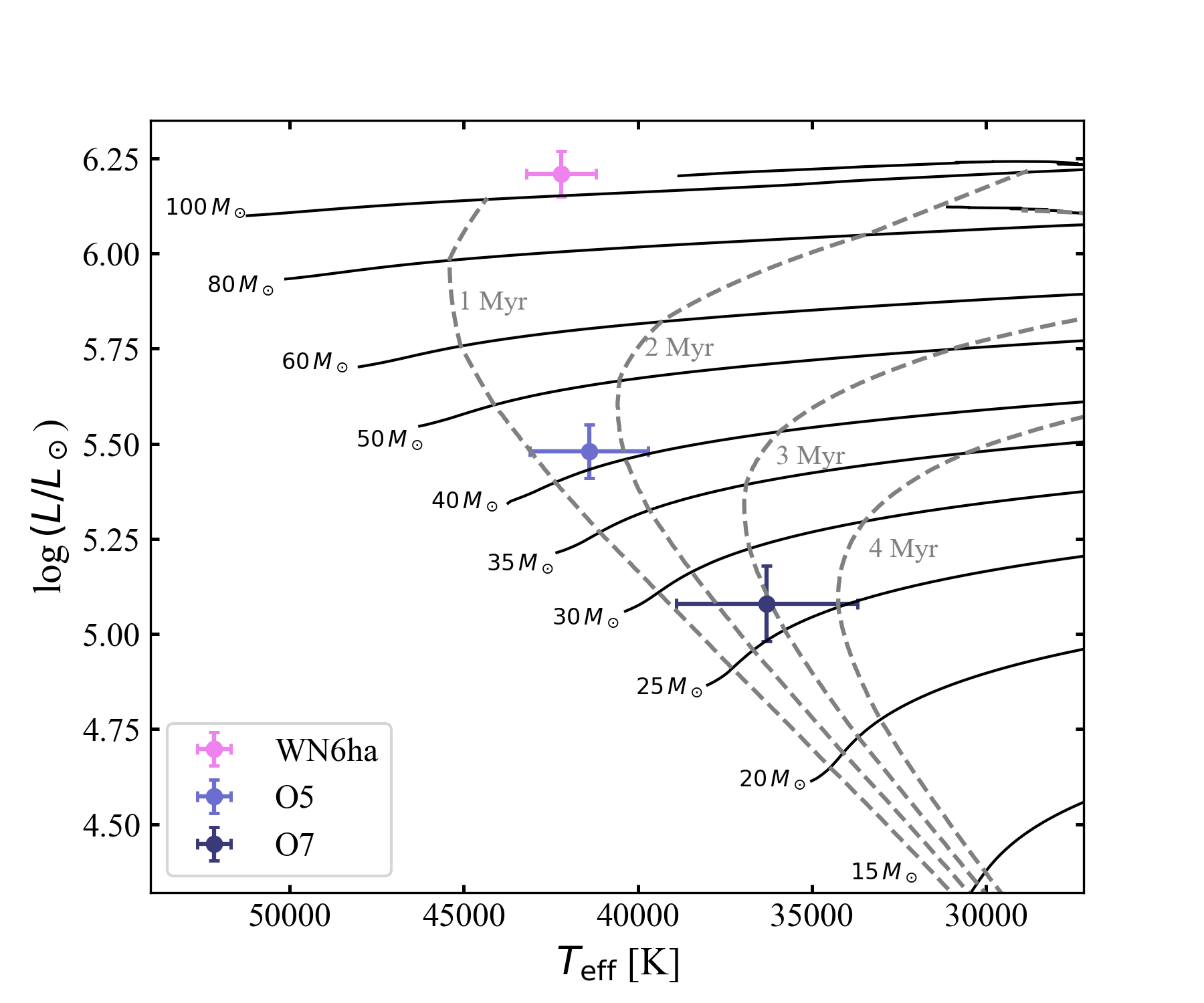}
    \caption{All three components in WR~25 shown on a Hertzprung-Russell diagram, with $T_{\rm eff}-{\rm log}\,L$ values taken from Table\,\ref{tab:parameters}. Also shown are evolutionary tracks computed with an initial rotational velocity of 150~km~s$^{-1}$ from \citet{2011Brott} for a few different masses (solid), and isochrones for a few different ages (dashed).}
    \label{fig:hrpos}
\end{figure}

The evolutionary masses for the primary and secondary are consistent with their dynamical masses within error bars, although the evolutionary masses do appear to be systematically higher. The dynamical masses are currently constrained with only one interferometric epoch; more epochs will allow for better mass measurements and hence better comparison with evolutionary masses.  Additionally, the primary and secondary seem to be coeval, and a significant discrepancy arises only when nitrogen surface abundances are included. The tertiary component remains compatible with coevality within $2\sigma$ uncertainties. Figure\,\ref{fig:hrpos} shows all three components on a Hertzprung-Russel diagram.


\subsection{The orbit of the tertiary}
\label{sec:ter_orb}

\begin{figure}
    \centering
    \includegraphics[trim={0cm 0cm 0cm 16.5cm}, clip, width=0.95\linewidth]{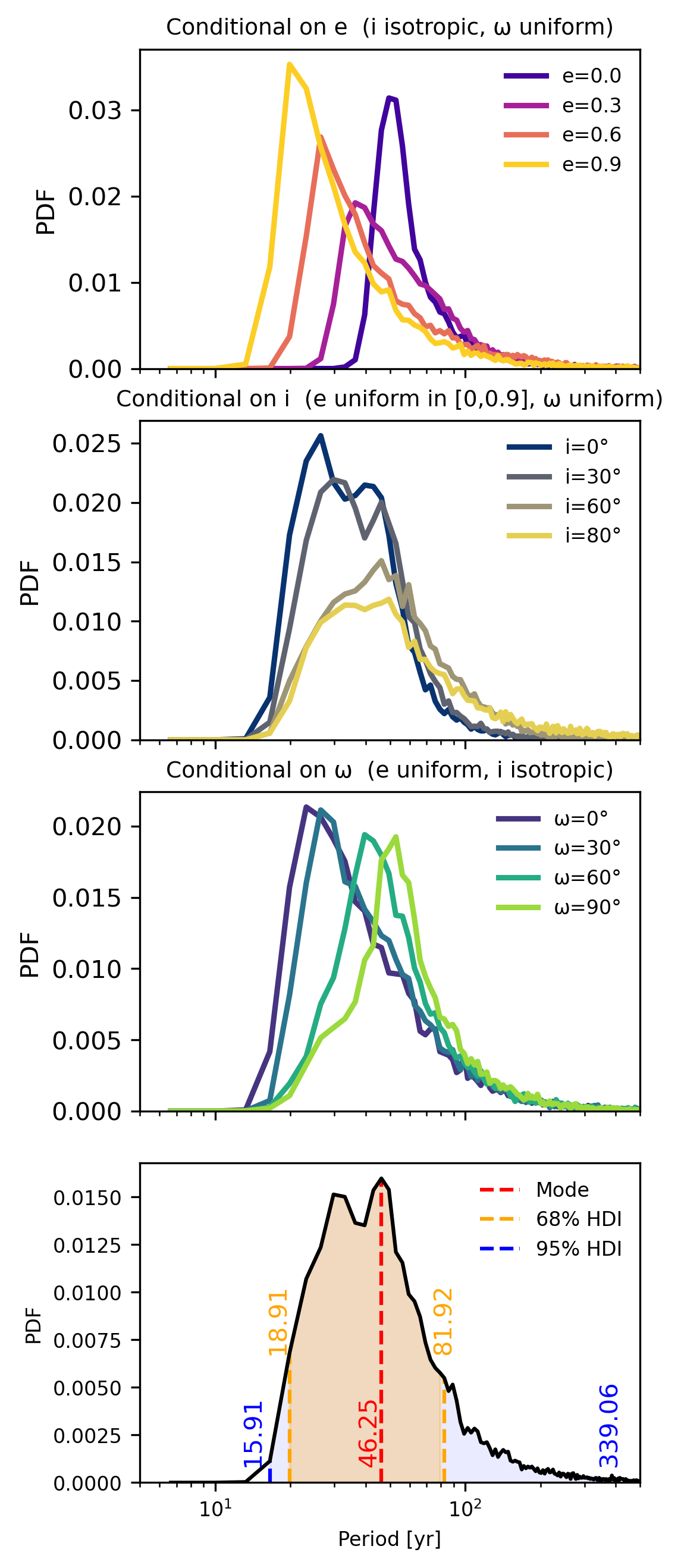}
    \caption{The probability distribution function (PDF) of the tertiary orbital period based on estimated masses and marginalized over random orientation of the system. Also shown are the mode (red), and the 68\% (orange) and 95\% (blue) highest density interval (HDI) regions.}
    \label{fig:P_ter}
\end{figure}

With a robust measurement of the relative astrometric position of the tertiary and an estimate of its mass, it is possible to broadly constrain its orbital properties. We used the methods described in \citet{2025Tramper} and Millour et al. (submitted) to estimate the posterior distribution of the orbital period/separation and velocities given the masses of $M_1+M_2$ (dynamical) and of $M_3$ (evolutionary), and the observed angular separations. We adopted the total binary mass obtained in Sect.\,\ref{sec:masstot}, along with the distance, $\Delta E$, and $\Delta N$. We marginalized over a random orientation of the system and assumed a flat eccentricity prior between 0.0 -- 0.9. The resulting probability distribution of tertiary periods is shown in Figure\,\ref{fig:P_ter}.

The most likely tertiary period (68\% highest density interval) ranges from 19 -- 82 yr (corresponding to a semi-major axis of 38 -- 94 au), while its sky-plane angular velocity ranges from 1.3 -- 3.2 mas yr$^{-1}$, which should be easily detectable in future VLTI observations. The expected RV semi-amplitude is only of a few ${\rm km\,s}^{-1}$ and will be much harder to detect, if at all possible.

\section{Conclusions} \label{sec:disc}
 
We present a VLTI/PIONIER interferometric observation of the massive colliding wind binary WR~25, spatially resolving the two spectroscopic components. We derive the total dynamical mass of the binary $M_{\rm tot}=93\pm18\,M_\odot$, and subsequently the component masses $M_1 = 62\pm13\,M_\odot$ and $M_2 = 31\pm7\,M_\odot$. More measurements of the relative positions of the binary components will be crucial to constrain the 3D binary orbit. We also detect the third ``intruder'' component reported by \citet{2026Gosset} at an angular separation of $27.69\pm0.02$ mas from the primary, establishing WR~25 as a likely hierarchical triple system.

We revisit spectroscopic data of WR~25 to perform spectral disentangling and separate spectra for all three components. Following detailed atmospheric analysis performed with CMFGEN for each component, we estimate their evolutionary masses and ages using single star evolution tracks in BONNSAI. All three components in WR~25 have evolutionary ages consistent within $2\sigma$, which are also consistent with the 1 -- 3 Myr age estimated for Trumpler 16 \citep{2012Hur}. 

Finally, we make predictions for the orbital properties of the tertiary component and estimate the outer orbital period to be in the range 19 -- 82 yr. The corresponding sky-plane angular velocity of the tertiary is expected to be 1 -- 3 mas yr$^{-1}$, meaning it is likely to show a significant shift in its position if observed now, as compared to 2019 when it was first observed with VLTI. More interferometric epochs can confirm the hierarchical triple nature of WR~25 beyond any doubt and also provide stronger constraints on the tertiary orbit.

WR~25 represents the rare case of a massive hierarchical triple where all three components are spatially resolved \citep[more common for B-type stars, see][]{2025Frost}. \citet{2025Sana} note that several O-type massive hierarchical triple systems have been identified with interferometry, but are often inadequately constrained due to the inner binary being too close to be spatially resolved \citep[e.g.][]{2026Bordier}. With all three components spatially resolved, WR~25 presents the opportunity to determine both the inner and outer orbits, and use their mutual orientations to study the formation and dynamical stability of the triple system \citep{2001Mardling,2004Tokovinin}. Additionally, measuring accurate and precise dynamical masses for all three components will be crucial to calibrate stellar physics for massive stars. For example, leveraging the component masses and coevality of the triple system, it will be possible to constrain internal mixing and wind mass-loss which carry significant uncertainties in current stellar evolution models. We therefore encourage continued interferometric observations of WR~25 to measure precise orbits for the inner binary, the tertiary, and the dynamical masses of all components.

\begin{acknowledgements}
We thank Vincent Bronner for sharing the extended BONNSAI evolutionary tracks and isochrones. This research is based on observations made with ESO Telescopes at the La Silla Paranal Observatory under program ID 0102.D-0234. This research is supported by the Flemish Government under the long-term structural Methusalem funding program, project SOUL: Stellar evolution in full glory, grant METH/24/012 at KU Leuven. TS acknowledges funding from the European Research Council (ERC) under the European Union’s Horizon 2020 and Horizon
Europe research and innovation programme (grant agreement number 101164755: METAL) and the Israel Science Foundation
(ISF) under grant number 0603225041. DP acknowledges financial support from the Research Foundation Flanders (FWO) under grant agreement No. 1256225N.

\end{acknowledgements}

\bibliographystyle{aa}
\bibliography{papers}

\begin{thebibliography}{66}
\expandafter\ifx\csname natexlab\endcsname\relax\def\natexlab#1{#1}\fi

\bibitem[{{Absil} {et~al.}(2011){Absil}, {Le Bouquin}, {Berger}, {Lagrange}, {Chauvin}, {Lazareff}, {Zins}, {Haguenauer}, {Jocou}, {Kern}, {Millan-Gabet}, {Rochat}, \& {Traub}}]{2011Absil}
{Absil}, O., {Le Bouquin}, J.~B., {Berger}, J.~P., {et~al.} 2011, \aap, 535, A68

\bibitem[{{Antokhin} {et~al.}(2008){Antokhin}, {Rauw}, {Vreux}, {van der Hucht}, \& {Brown}}]{2008Antokhin}
{Antokhin}, I.~I., {Rauw}, G., {Vreux}, J.~M., {van der Hucht}, K.~A., \& {Brown}, J.~C. 2008, \aap, 477, 593

\bibitem[{{Arora} {et~al.}(2019){Arora}, {Pandey}, \& {De Becker}}]{2019Arora}
{Arora}, B., {Pandey}, J.~C., \& {De Becker}, M. 2019, \mnras, 487, 2624

\bibitem[{{Barb{\'a}} {et~al.}(2022){Barb{\'a}}, {Gamen}, {Mart{\'\i}n-Ravelo}, {Arias}, \& {Morrell}}]{2022Barba}
{Barb{\'a}}, R.~H., {Gamen}, R.~C., {Mart{\'\i}n-Ravelo}, P., {Arias}, J.~I., \& {Morrell}, N.~I. 2022, \mnras, 516, 1149

\bibitem[{{Bonanos} {et~al.}(2004){Bonanos}, {Stanek}, {Udalski}, {Wyrzykowski}, {{\.Z}ebru{\'n}}, {Kubiak}, {Szyma{\'n}ski}, {Szewczyk}, {Pietrzy{\'n}ski}, \& {Soszy{\'n}ski}}]{2004Bonanos}
{Bonanos}, A.~Z., {Stanek}, K.~Z., {Udalski}, A., {et~al.} 2004, \apjl, 611, L33

\bibitem[{{Bordier} {et~al.}(2026){Bordier}, {Sana}, {Frost}, {Libert}, {Vrancken}, {Mahy}, {Toonen}, {Tramper}, {de Koter}, {Lacour}, {Le Bouquin}, \& {de Wit}}]{2026Bordier}
{Bordier}, E., {Sana}, H., {Frost}, A.~J., {et~al.} 2026, arXiv e-prints, arXiv:2606.28159

\bibitem[{{Brott} {et~al.}(2011){Brott}, {de Mink}, {Cantiello}, {Langer}, {de Koter}, {Evans}, {Hunter}, {Trundle}, \& {Vink}}]{2011Brott}
{Brott}, I., {de Mink}, S.~E., {Cantiello}, M., {et~al.} 2011, \aap, 530, A115

\bibitem[{{Crowther}(2007)}]{2007Crowther}
{Crowther}, P.~A. 2007, \araa, 45, 177

\bibitem[{{Crowther} {et~al.}(1995){Crowther}, {Smith}, {Hillier}, \& {Schmutz}}]{crowther95}
{Crowther}, P.~A., {Smith}, L.~J., {Hillier}, D.~J., \& {Schmutz}, W. 1995, \aap, 293, 427

\bibitem[{{de Mink} {et~al.}(2013){de Mink}, {Langer}, {Izzard}, {Sana}, \& {de Koter}}]{2013deMink}
{de Mink}, S.~E., {Langer}, N., {Izzard}, R.~G., {Sana}, H., \& {de Koter}, A. 2013, \apj, 764, 166

\bibitem[{{Deshmukh} {et~al.}(2024){Deshmukh}, {Sana}, {M{\'e}rand}, {Bordier}, {Langer}, {Bodensteiner}, {Dsilva}, {Frost}, {Gosset}, {Le Bouquin}, {Lefever}, {Mahy}, {Patrick}, {Reggiani}, {Sander}, {Shenar}, {Tramper}, {Villase{\~n}or}, \& {Waisberg}}]{2024Deshmukh}
{Deshmukh}, K., {Sana}, H., {M{\'e}rand}, A., {et~al.} 2024, \aap, 692, A109

\bibitem[{{Deshmukh} {et~al.}(2025){Deshmukh}, {Shenar}, {M{\'e}rand}, {Sana}, {Marchant}, {Wade}, {Bodensteiner}, {Chen{\'e}}, {Frost}, {Gilkis}, {Langer}, \& {Oskinova}}]{2025Deshmukh}
{Deshmukh}, K., {Shenar}, T., {M{\'e}rand}, A., {et~al.} 2025, \aap, 695, L20

\bibitem[{{Eldridge} \& {Stanway}(2022)}]{2022Eldridge}
{Eldridge}, J.~J. \& {Stanway}, E.~R. 2022, \araa, 60, 455

\bibitem[{{Frost} {et~al.}(2025){Frost}, {Sana}, {Le Bouquin}, {Perets}, {Bodensteiner}, {Igoshev}, {Banyard}, {Mahy}, {M{\'e}rand}, \& {Ram{\'\i}rez-Agudelo}}]{2025Frost}
{Frost}, A.~J., {Sana}, H., {Le Bouquin}, J.-B., {et~al.} 2025, \aap, 701, A171

\bibitem[{{Frost} {et~al.}(2024){Frost}, {Sana}, {Mahy}, {Wade}, {Barron}, {Le Bouquin}, {M{\'e}rand}, {Schneider}, {Shenar}, {Barb{\'a}}, {Bowman}, {Fabry}, {Farhang}, {Marchant}, {Morrell}, \& {Smoker}}]{2024Frost}
{Frost}, A.~J., {Sana}, H., {Mahy}, L., {et~al.} 2024, Science, 384, 214

\bibitem[{{Gallenne} {et~al.}(2015){Gallenne}, {M{\'e}rand}, {Kervella}, {Monnier}, {Schaefer}, {Baron}, {Breitfelder}, {Le Bouquin}, {Roettenbacher}, {Gieren}, {Pietrzy{\'n}ski}, {McAlister}, {ten Brummelaar}, {Sturmann}, {Sturmann}, {Turner}, {Ridgway}, \& {Kraus}}]{2015Gallenne}
{Gallenne}, A., {M{\'e}rand}, A., {Kervella}, P., {et~al.} 2015, \aap, 579, A68

\bibitem[{{Gamen} {et~al.}(2006){Gamen}, {Gosset}, {Morrell}, {Niemela}, {Sana}, {Naz{\'e}}, {Rauw}, {Barb{\'a}}, \& {Solivella}}]{2006Gamen}
{Gamen}, R., {Gosset}, E., {Morrell}, N., {et~al.} 2006, \aap, 460, 777

\bibitem[{{Gosset} {et~al.}(2026){Gosset}, {Gamen}, {Mahy}, {Morrell}, {Sana}, \& {Barb\'a}}]{2026Gosset}
{Gosset}, E., {Gamen}, R., {Mahy}, L., {et~al.} 2026, Bull. Soc. R. Sci. Li\`ege, in press

\bibitem[{{Grevesse} {et~al.}(2010){Grevesse}, {Asplund}, {Sauval}, \& {Scott}}]{Grevesse2010}
{Grevesse}, N., {Asplund}, M., {Sauval}, A.~J., \& {Scott}, P. 2010, \apss, 328, 179

\bibitem[{{Hadrava}(1995)}]{Hadrava1995}
{Hadrava}, P. 1995, \aaps, 114, 393

\bibitem[{{Hawcroft} {et~al.}(2024){Hawcroft}, {Sana}, {Mahy}, {Sundqvist}, {de Koter}, {Crowther}, {Bestenlehner}, {Brands}, {David-Uraz}, {Decin}, {Erba}, {Garcia}, {Hamann}, {Herrero}, {Ignace}, {Kee}, {Kub{\'a}tov{\'a}}, {Lefever}, {Moffat}, {Najarro}, {Oskinova}, {Pauli}, {Prinja}, {Puls}, {Sander}, {Shenar}, {St-Louis}, {ud-Doula}, \& {Vink}}]{hawcroft2024}
{Hawcroft}, C., {Sana}, H., {Mahy}, L., {et~al.} 2024, \aap, 688, A105

\bibitem[{{Hillier} \& {Miller}(1998)}]{Hillier1998}
{Hillier}, D.~J. \& {Miller}, D.~L. 1998, \apj, 496, 407

\bibitem[{{Holdsworth} {et~al.}(2024){Holdsworth}, {Richardson}, {Schaefer}, {Eldridge}, {Hill}, {Spejcher}, {Mackey}, {Moffat}, {Navarete}, {Monnier}, {Kraus}, {Le Bouquin}, {Anugu}, {Chhabra}, {Codron}, {Ennis}, {Gardner}, {Gutierrez}, {Ibrahim}, {Labdon}, {Lanthermann}, \& {Setterholm}}]{2024Holdsworth}
{Holdsworth}, A., {Richardson}, N., {Schaefer}, G.~H., {et~al.} 2024, \apj, 977, 185

\bibitem[{{Hur} {et~al.}(2012){Hur}, {Sung}, \& {Bessell}}]{2012Hur}
{Hur}, H., {Sung}, H., \& {Bessell}, M.~S. 2012, \aj, 143, 41

\bibitem[{{Ilijic} {et~al.}(2004){Ilijic}, {Hensberge}, {Pavlovski}, \& {Freyhammer}}]{Ilijic2004}
{Ilijic}, S., {Hensberge}, H., {Pavlovski}, K., \& {Freyhammer}, L.~M. 2004, in Astronomical Society of the Pacific Conference Series, Vol. 318, Spectroscopically and Spatially Resolving the Components of the Close Binary Stars, ed. R.~W. {Hilditch}, H.~{Hensberge}, \& K.~{Pavlovski}, 111--113

\bibitem[{{Kaufer} {et~al.}(1999){Kaufer}, {Stahl}, {Tubbesing}, {N{\o}rregaard}, {Avila}, {Francois}, {Pasquini}, \& {Pizzella}}]{1999Kaufer}
{Kaufer}, A., {Stahl}, O., {Tubbesing}, S., {et~al.} 1999, The Messenger, 95, 8

\bibitem[{{Kruckow} {et~al.}(2018){Kruckow}, {Tauris}, {Langer}, {Kramer}, \& {Izzard}}]{2018Kruckow}
{Kruckow}, M.~U., {Tauris}, T.~M., {Langer}, N., {Kramer}, M., \& {Izzard}, R.~G. 2018, \mnras, 481, 1908

\bibitem[{{Lamontagne} {et~al.}(1996){Lamontagne}, {Moffat}, {Drissen}, {Robert}, \& {Matthews}}]{1996Lamontagne}
{Lamontagne}, R., {Moffat}, A. F.~J., {Drissen}, L., {Robert}, C., \& {Matthews}, J.~M. 1996, \aj, 112, 2227

\bibitem[{{Langer}(2012)}]{2012Langer}
{Langer}, N. 2012, \araa, 50, 107

\bibitem[{{Langer} {et~al.}(1994){Langer}, {Hamann}, {Lennon}, {Najarro}, {Pauldrach}, \& {Puls}}]{1994Langer}
{Langer}, N., {Hamann}, W.~R., {Lennon}, M., {et~al.} 1994, \aap, 290, 819

\bibitem[{{Lanthermann} {et~al.}(2023){Lanthermann}, {Le Bouquin}, {Sana}, {M{\'e}rand}, {Monnier}, {Perraut}, {Frost}, {Mahy}, {Gosset}, {De Becker}, {Kraus}, {Anugu}, {Davies}, {Ennis}, {Gardner}, {Labdon}, {Setterholm}, {ten Brummelaar}, \& {Schaefer}}]{2023Lanthermann}
{Lanthermann}, C., {Le Bouquin}, J.-B., {Sana}, H., {et~al.} 2023, \aap, 672, A6

\bibitem[{{Le Bouquin} {et~al.}(2011){Le Bouquin}, {Berger}, {Lazareff}, {Zins}, {Haguenauer}, {Jocou}, {Kern}, {Millan-Gabet}, {Traub}, {Absil}, {Augereau}, {Benisty}, {Blind}, {Bonfils}, {Bourget}, {Delboulbe}, {Feautrier}, {Germain}, {Gitton}, {Gillier}, {Kiekebusch}, {Kluska}, {Knudstrup}, {Labeye}, {Lizon}, {Monin}, {Magnard}, {Malbet}, {Maurel}, {M{\'e}nard}, {Micallef}, {Michaud}, {Montagnier}, {Morel}, {Moulin}, {Perraut}, {Popovic}, {Rabou}, {Rochat}, {Rojas}, {Roussel}, {Roux}, {Stadler}, {Stefl}, {Tatulli}, \& {Ventura}}]{2011LeBouquin}
{Le Bouquin}, J.~B., {Berger}, J.~P., {Lazareff}, B., {et~al.} 2011, \aap, 535, A67

\bibitem[{{Le Bouquin} {et~al.}(2017){Le Bouquin}, {Sana}, {Gosset}, {De Becker}, {Duvert}, {Absil}, {Anthonioz}, {Berger}, {Ertel}, {Grellmann}, {Guieu}, {Kervella}, {Rabus}, \& {Willson}}]{2017LeBouquin}
{Le Bouquin}, J.-B., {Sana}, H., {Gosset}, E., {et~al.} 2017, \aap, 601, A34

\bibitem[{{Mahy} {et~al.}(2017){Mahy}, {Damerdji}, {Gosset}, {Nitschelm}, {Eenens}, {Sana}, \& {Klotz}}]{Mahy2017}
{Mahy}, L., {Damerdji}, Y., {Gosset}, E., {et~al.} 2017, \aap, 607, A96

\bibitem[{{Marchant} \& {Bodensteiner}(2024)}]{2024Marchant}
{Marchant}, P. \& {Bodensteiner}, J. 2024, \araa, 62, 21

\bibitem[{{Mardling} \& {Aarseth}(2001)}]{2001Mardling}
{Mardling}, R.~A. \& {Aarseth}, S.~J. 2001, \mnras, 321, 398

\bibitem[{{Martins} \& {Plez}(2006)}]{2006Martins}
{Martins}, F. \& {Plez}, B. 2006, \aap, 457, 637

\bibitem[{{Massey} {et~al.}(2025){Massey}, {Bodansky}, {Penny}, {Morrell}, \& {Neugent}}]{2025Massey}
{Massey}, P., {Bodansky}, S., {Penny}, L.~R., {Morrell}, N.~I., \& {Neugent}, K.~F. 2025, \apj, 990, 52

\bibitem[{{M{\'e}rand}(2022)}]{2022Merand}
{M{\'e}rand}, A. 2022, in Society of Photo-Optical Instrumentation Engineers (SPIE) Conference Series, Vol. 12183, Optical and Infrared Interferometry and Imaging VIII, ed. A.~{M{\'e}rand}, S.~{Sallum}, \& J.~{Sanchez-Bermudez}, 121831N

\bibitem[{{Pandey} {et~al.}(2014){Pandey}, {Pandey}, \& {Karmakar}}]{2014Pandey}
{Pandey}, J.~C., {Pandey}, S.~B., \& {Karmakar}, S. 2014, \apj, 788, 84

\bibitem[{{Pollock} \& {Corcoran}(2006)}]{2006Pollock}
{Pollock}, A.~M.~T. \& {Corcoran}, M.~F. 2006, \aap, 445, 1093

\bibitem[{{Pradhan} {et~al.}(2021){Pradhan}, {Huenemoerder}, {Ignace}, {Pollock}, \& {Nichols}}]{2021Pradhan}
{Pradhan}, P., {Huenemoerder}, D.~P., {Ignace}, R., {Pollock}, A.~M.~T., \& {Nichols}, J.~S. 2021, \apj, 915, 114

\bibitem[{{Raassen} {et~al.}(2003){Raassen}, {van der Hucht}, {Mewe}, {Antokhin}, {Rauw}, {Vreux}, {Schmutz}, \& {G{\"u}del}}]{2003Raassen}
{Raassen}, A.~J.~J., {van der Hucht}, K.~A., {Mewe}, R., {et~al.} 2003, \aap, 402, 653

\bibitem[{{Rainot} {et~al.}(2020){Rainot}, {Reggiani}, {Sana}, {Bodensteiner}, {Gomez-Gonzalez}, {Absil}, {Christiaens}, {Delorme}, {Almeida}, {Caballero-Nieves}, {De Ridder}, {Kratter}, {Lacour}, {Le Bouquin}, {Pueyo}, \& {Zinnecker}}]{2020Rainot}
{Rainot}, A., {Reggiani}, M., {Sana}, H., {et~al.} 2020, \aap, 640, A15

\bibitem[{{Rauw} {et~al.}(2004){Rauw}, {De Becker}, {Naz{\'e}}, {Crowther}, {Gosset}, {Sana}, {van der Hucht}, {Vreux}, \& {Williams}}]{2004Rauw}
{Rauw}, G., {De Becker}, M., {Naz{\'e}}, Y., {et~al.} 2004, \aap, 420, L9

\bibitem[{{Richardson} {et~al.}(2024){Richardson}, {Schaefer}, {Eldridge}, {Spejcher}, {Holdsworth}, {Lau}, {Monnier}, {Moffat}, {Weigelt}, {Williams}, {Kraus}, {Le Bouquin}, {Anugu}, {Chhabra}, {Codron}, {Ennis}, {Gardner}, {Gutierrez}, {Ibrahim}, {Labdon}, {Lanthermann}, \& {Setterholm}}]{2024Richardson}
{Richardson}, N.~D., {Schaefer}, G.~H., {Eldridge}, J.~J., {et~al.} 2024, \apj, 977, 78

\bibitem[{{Sana} {et~al.}(2025){Sana}, {Bordier}, {Deshmukh}, {Frost}, {Keskar}, {Lanthermann}, {Lefever}, {Mahy}, {Sander}, {Shenar}, \& {Tramper}}]{2025Sana}
{Sana}, H., {Bordier}, E., {Deshmukh}, K., {et~al.} 2025, arXiv e-prints, arXiv:2512.00444

\bibitem[{{Sana} {et~al.}(2012){Sana}, {de Mink}, {de Koter}, {Langer}, {Evans}, {Gieles}, {Gosset}, {Izzard}, {Le Bouquin}, \& {Schneider}}]{2012Sana}
{Sana}, H., {de Mink}, S.~E., {de Koter}, A., {et~al.} 2012, Science, 337, 444

\bibitem[{{Sana} {et~al.}(2014){Sana}, {Le Bouquin}, {Lacour}, {Berger}, {Duvert}, {Gauchet}, {Norris}, {Olofsson}, {Pickel}, {Zins}, {Absil}, {de Koter}, {Kratter}, {Schnurr}, \& {Zinnecker}}]{2014Sana}
{Sana}, H., {Le Bouquin}, J.~B., {Lacour}, S., {et~al.} 2014, \apjs, 215, 15

\bibitem[{{Sasaki} {et~al.}(2024){Sasaki}, {Robrade}, {Krause}, {Knies}, {Tsuge}, {P{\"u}hlhofer}, \& {Strong}}]{2024Sasaki}
{Sasaki}, M., {Robrade}, J., {Krause}, M. G.~H., {et~al.} 2024, \aap, 682, A172

\bibitem[{{Schneider} {et~al.}(2014){Schneider}, {Langer}, {de Koter}, {Brott}, {Izzard}, \& {Lau}}]{Schneider2014}
{Schneider}, F.~R.~N., {Langer}, N., {de Koter}, A., {et~al.} 2014, \aap, 570, A66

\bibitem[{{Shenar} {et~al.}(2017){Shenar}, {Richardson}, {Sablowski}, {Hainich}, {Sana}, {Moffat}, {Todt}, {Hamann}, {Oskinova}, {Sander}, {Tramper}, {Langer}, {Bonanos}, {de Mink}, {Gr{\"a}fener}, {Crowther}, {Vink}, {Almeida}, {de Koter}, {Barb{\'a}}, {Herrero}, \& {Ulaczyk}}]{2017Shenar}
{Shenar}, T., {Richardson}, N.~D., {Sablowski}, D.~P., {et~al.} 2017, \aap, 598, A85

\bibitem[{{Shenar} {et~al.}(2021){Shenar}, {Sana}, {Marchant}, {Pablo}, {Richardson}, {Moffat}, {Van Reeth}, {Barb{\'a}}, {Bowman}, {Broos}, {Crowther}, {Clark}, {de Koter}, {de Mink}, {Dsilva}, {Gr{\"a}fener}, {Howarth}, {Langer}, {Mahy}, {Ma{\'\i}z Apell{\'a}niz}, {Pollock}, {Schneider}, {Townsley}, \& {Vink}}]{2021Shenar}
{Shenar}, T., {Sana}, H., {Marchant}, P., {et~al.} 2021, \aap, 650, A147

\bibitem[{{Shull} {et~al.}(2021){Shull}, {Darling}, \& {Danforth}}]{2021Shull}
{Shull}, J.~M., {Darling}, J., \& {Danforth}, C.~W. 2021, \apj, 914, 18

\bibitem[{{Sim{\'o}n-D{\'\i}az} \& {Herrero}(2007)}]{Simon-Diaz2007}
{Sim{\'o}n-D{\'\i}az}, S. \& {Herrero}, A. 2007, \aap, 468, 1063

\bibitem[{{Sim{\'o}n-D{\'\i}az} \& {Herrero}(2014)}]{Simon-Diaz2014}
{Sim{\'o}n-D{\'\i}az}, S. \& {Herrero}, A. 2014, \aap, 562, A135

\bibitem[{{Skrutskie} {et~al.}(2003){Skrutskie}, {Cutri}, {Stiening}, {Weinberg}, {Schneider}, {Carpenter}, {Beichman}, {Capps}, {Chester}, {Elias}, {Huchra}, {Liebert}, {Lonsdale}, {Monet}, {Price}, {Seitzer}, {Jarrett}, {Kirkpatrick}, {Gizis}, {Howard}, {Evans}, {Fowler}, {Fullmer}, {Hurt}, {Light}, {Kopan}, {Marsh}, {McCallon}, {Tam}, {Van Dyk}, \& {Wheelock}}]{2003Skrutskie}
{Skrutskie}, M.~F., {Cutri}, R.~M., {Stiening}, R., {et~al.} 2003, {2MASS All-Sky Point Source Catalog}, NASA IPAC DataSet, IRSA2

\bibitem[{{Smith} \& {Conti}(2008)}]{2008Smith}
{Smith}, N. \& {Conti}, P.~S. 2008, \apj, 679, 1467

\bibitem[{{St.-Louis} {et~al.}(1988){St.-Louis}, {Moffat}, {Drissen}, {Bastien}, \& {Robert}}]{1988StLouis}
{St.-Louis}, N., {Moffat}, A. F.~J., {Drissen}, L., {Bastien}, P., \& {Robert}, C. 1988, \apj, 330, 286

\bibitem[{{Stegmann} {et~al.}(2026){Stegmann}, {Antonini}, {Olejak}, {Biscoveanu}, {Raymond}, {Rinaldi}, \& {Flanagan}}]{2026Stegmann}
{Stegmann}, J., {Antonini}, F., {Olejak}, A., {et~al.} 2026, \apjl, 1000, L59

\bibitem[{{The LIGO Scientific Collaboration} {et~al.}(2025){The LIGO Scientific Collaboration}, {the Virgo Collaboration}, {the KAGRA Collaboration}, {Abac}, {Abouelfettouh}, {Acernese}, {Ackley}, {Adamcewicz}, {Adhicary}, {Adhikari}, {Adhikari}, {Adhikari}, {Adkins}, {Afroz}, {Agapito}, {Agarwal}, {Agathos}, {Aggarwal}, {Aggarwal}, {Aguiar}, {Ahrend}, {Aiello}, {Ain}, {Ajith}, {Akutsu}, {Albanesi}, {Ali}, {Al-Kershi}, {All{\'e}n{\'e}}, {Allocca}, {Al-Shammari}, {Altin}, {Alvarez-Lopez}, {Amar}, {Amarasinghe}, {Amato}, {Amicucci}, {Amra}, {Ananyeva}, {Anderson}, {Anderson}, {Andia}, {Ando}, {Andr{\'e}s-Carcasona}, {Andri{\'c}}, {Anglin}, {Ansoldi}, {Antelis}, {Antier}, {Aoumi}, {Appavuravther}, {Appert}, {Apple}, {Arai}, {Araya}, {Araya}, {Arca Sedda}, {Areeda}, {Aritomi}, {Armato}, {Armstrong}, {Arnaud}, {Arogeti}, {Aronson}, {Arun}, {Ashton}, {Aso}, {Asprea}, {Assiduo}, {Assis de Souza Melo}, {Aston}, {Astone}, {Attadio}, {Aubin}, {AultONeal}, {Avallone}, {Avila}, {Babak}, {Badger}, {Bae}, {Bagnasco},
  {Baiotti}, {Bajpai}, {Baka}, {Baker}, {Baker}, {Baker}, {Baldi}, {Baldicchi}, {Ball}, {Ballardin}, {Ballmer}, {Banagiri}, {Banerjee}, {Bankar}, {Baptiste}, {Baral}, {Baratti}, {Barayoga}, {Barish}, {Barker}, {Barman}, {Barneo}, {Barone}, {Barr}, {Barsotti}, {Barsuglia}, {Barta}, {Bartoletti}, {Barton}, {Bartos}, {Basalaev}, {Bassiri}, {Basti}, {Bawaj}, {Baxi}, {Bayley}, {Baylor}, {Baynard}, {Bazzan}, {Bedakihale}, {Beirnaert}, {Bejger}, {Belardinelli}, {Bell}, {Bellie}, {Bellizzi}, {Benoit}, {Bentara}, {Bentley}, {Ben Yaala}, {Bera}, {Bergamin}, {Berger}, {Bernuzzi}, {Beroiz}, {Berry}, {Bersanetti}, {Bertheas}, {Bertolini}, {Betzwieser}, {Beveridge}, {Bevilacqua}, {Bevins}, {Bhandare}, {Bhatt}, {Bhattacharjee}, {Bhattacharyya}, {Bhaumik}, {Biancalana}, {Bianchi}, {Bilenko}, {Billingsley}, {Binetti}, {Bini}, {Binu}, {Biot}, {Birnholtz}, {Biscoveanu}, {Bisht}, {Bitossi}, {Bizouard}, {Blaber}, {Blackburn}, {Blagg}, {Blair}, {Blair}, {Bode}, {Boettner}, {Boileau}, {Boldrini}, {Bolingbroke}, {Bolliand},
  {Bonavena}, {Bondarescu}, {Bondu}, {Bonilla}, {Bonilla}, {Bonino}, {Bonnand}, {Borchers}, {Borhanian}, {Boschi}, {Bose}, {Bossilkov}, {Bothra}, {Boudon}, {Bourg}, {Boyle}, {Bozzi}, {Bradaschia}, {Brady}, {Branch}, {Branchesi}, {Braun}, {Briant}, {Brillet}, {Brinkmann}, {Brockill}, \& {Brockmueller}}]{2025gwtc}
{The LIGO Scientific Collaboration}, {the Virgo Collaboration}, {the KAGRA Collaboration}, {et~al.} 2025, arXiv e-prints, arXiv:2508.18082

\bibitem[{{Tokovinin}(2004)}]{2004Tokovinin}
{Tokovinin}, A. 2004, in Revista Mexicana de Astronomia y Astrofisica Conference Series, Vol.~21, Revista Mexicana de Astronomia y Astrofisica Conference Series, ed. C.~{Allen} \& C.~{Scarfe}, 7--14

\bibitem[{{Tramper} {et~al.}(2025){Tramper}, {Sana}, {de Koter}, \& {Pauwels}}]{2025Tramper}
{Tramper}, F., {Sana}, H., {de Koter}, A., \& {Pauwels}, T. 2025, arXiv e-prints, arXiv:2509.18431

\bibitem[{{Tramper} {et~al.}(2016){Tramper}, {Sana}, {Fitzsimons}, {de Koter}, {Kaper}, {Mahy}, \& {Moffat}}]{2016Tramper}
{Tramper}, F., {Sana}, H., {Fitzsimons}, N.~E., {et~al.} 2016, \mnras, 455, 1275

\bibitem[{{Walborn}(1972)}]{Walborn1972}
{Walborn}, N.~R. 1972, \aj, 77, 312

\bibitem[{{Woosley}(2019)}]{2019Woosley}
{Woosley}, S.~E. 2019, \apj, 878, 49

\end{thebibliography}

\end{document}